\documentclass[journal]{IEEEtran}
\usepackage{amsmath,amsfonts,amsthm}
\usepackage{array}
\usepackage[caption=false,font=footnotesize,labelfont=rm,textfont=rm]{subfig}
\usepackage{textcomp}
\usepackage{stfloats}
\usepackage{url}
\usepackage{verbatim}
\usepackage{graphicx}
\usepackage{amssymb}
\usepackage{bm}
\usepackage{mathrsfs}
\usepackage{graphicx} 
\usepackage{subfig}
\usepackage{booktabs}
\usepackage{diagbox}
\usepackage {hyperref}
\usepackage{utfsym}
\usepackage{stmaryrd}
\usepackage{multirow}
\usepackage{makecell}
\usepackage{threeparttable}
\usepackage[ruled,linesnumbered,vlined]{algorithm2e}
\begin{document}
	
\title{A Spatial-Domain Coordinated Control Method for CAVs at Unsignalized Intersections Considering Motion Uncertainty}
	
\author{Tong Zhao, Nikolce Murgovski, Baigen Cai, \IEEEmembership{Senior Member, IEEE}, and Wei ShangGuan, \IEEEmembership{Member, IEEE}
	
\thanks{This work was supported in part by the Beijing Outstanding Young Scientist Program under Grant JWZQ20240101010; in part by the Joint Fund of the Ministry of Education for Equipment Pre-Research under Grant 8091B022238; and in part by the China Scholarship Council under Grant 202307090098. \emph{(Corresponding author: Wei ShangGuan.)}}
	
\thanks{Tong Zhao is with the School of Automation and Intelligence, Beijing Jiaotong University, Beijing 100044, China (e-mail: tong\_zhao@bjtu.edu.cn).}
	
\thanks{Nikolce Murgovski is with the Department of Electrical Engineering, Chalmers University of Technology, 412 96 Gothenburg, Sweden (e-mail: nikolce.murgovski@chalmers.se).}
	
\thanks{Baigen Cai and Wei ShangGuan are with the School of Automation and Intelligence, the State Key Laboratory of Advanced Rail Autonomous Operation, and the Research Center for Autonomous Intelligence and Unmanned Systems, Beijing Jiaotong University, Beijing 100044, China (e-mail: bgcai@bjtu.edu.cn; wshg@bjtu.edu.cn).}}
	
\markboth{}
{}
	
\IEEEpubid{}
	
\maketitle
	
\begin{abstract}
	
Coordinated control of connected and automated vehicles (CAVs) emerges as a promising technology to improve traffic safety, efficiency, and sustainability. Meanwhile, mixed traffic, where CAVs coexist with conventional human-driven vehicles (HDVs), represents an upcoming and necessary stage in the development of intelligent transportation systems. Considering the motion uncertainty of HDVs, this paper proposes a coordinated control method for trajectory planning of CAVs at an unsignalized intersection in mixed traffic. By sampling in distance and using an exact change of variables, the coordinated control problem is formulated in the spatial domain as a nonlinear program, thereby allowing for unified linear collision avoidance constraints to handle vehicle crossing, following, merging, and diverging conflicts. The motion uncertainty of HDVs is decoupled and modeled as path uncertainty and speed uncertainty, whereby the robustness of collision avoidance is ensured in both spatial and temporal dimensions. The prediction deviation for HDVs is compensated by receding horizon optimization, and a real-time iteration (RTI) scheme is developed to improve computational efficiency. Simulation case studies are conducted to validate the efficacy, robustness, and potential for real-time application of the proposed methods. The results show that the proposed control scheme provides collision-free and smooth trajectories with state and control constraints satisfied. Compared with the converged baseline, the RTI scheme reduces the computation time by orders of magnitude, and the solution deviation is less than 2.3\%, demonstrating a favorable trade-off between computational effort and optimality.

\end{abstract}
	
	
\section{Introduction}

\subsection{Motivation}

\IEEEPARstart{C}{onnected} and automated vehicles (CAVs) are considered one of the disruptive technologies that are expected to change the transportation landscape \cite{matin2023impacts}, \cite{rahman2023impacts}. With the aid of vehicle-to-everything (V2X) communication, CAVs are able to achieve cooperative driving, which holds great potential for improving traffic safety, efficiency, and sustainability \cite{zhang2023review}, \cite{yangjie2024towards}. 

Unsignalized intersections, being common and prone to crashes and congestion \cite{zhang2011unsignalized}, \cite{mousavi2021investigating}, have been widely regarded as typical application scenarios for cooperative driving \cite{li2024theoretical}. Various methods and techniques have been proposed for cooperative decision-making and control of CAVs at unsignalized intersections \cite{bejarbaneh2024exploring}, \cite{zhong2021autonomous}, \cite{rios2017survey}, \cite{chen2016cooperative}. However, most of them only work for fully cooperative systems with $100\,$\% CAV penetration. Over a long period of time, CAVs and conventional human-driven vehicles (HDVs) will coexist and thus constitute mixed traffic, which is an inevitable stage in the evolution of intelligent transportation systems towards full autonomy. Therefore, it is significant to consider the motion uncertainty of HDVs and investigate coordinated control methods for CAVs applicable to unsignalized intersections in mixed traffic.

\subsection{Literature Review}

From the perspective of research ideas, existing approaches to cooperative driving at unsignalized intersections can be classified into two categories: 1) one-stage approaches, which directly solve for the trajectories of CAVs; 2) two-stage approaches, which decouple the cooperative driving problem into two parts: crossing scheduling and trajectory planning, and first schedule the crossing order of all vehicles, and then plan the trajectories of CAVs based on this. The one-stage approach tends to embed complex safety constraints, resulting in a highly nonlinear and nonconvex optimization problem that is difficult to solve \cite{kamal2015vehicle}, \cite{ahn2020abstraction}, \cite{luo2023real}. In contrast, the two-stage approach reduces the problem complexity through decoupling and provides the flexibility that each stage can be handled by different methods. In the following, we focus on the research progress of the two-stage approach.

In terms of crossing scheduling, some strategies determine the crossing order based on heuristic rules, e.g., resource reservation \cite{dresner2008multiagent}, priority assignment \cite{alonso2011autonomous}, clearing policy \cite{ahmane2013modeling}. Such schemes can be implemented quickly online, but their solutions are not guaranteed to be optimal or good enough. Another class of strategies solves for the optimal crossing order by constructing an optimization problem, typically a mixed-integer linear program \cite{muller2016time}, \cite{fayazi2017optimal}. However, the computational complexity increases dramatically with the problem size, making these strategies intractable for real-time applications. To balance the computational effort and the optimality, heuristic optimization scheduling algorithms based on Monte Carlo tree search have been developed \cite{xu2020cooperative}, \cite{luo2024computationally}, \cite{xue2025conflict}. In addition, deep reinforcement learning techniques have been utilized to improve scheduling adaptability \cite{lombard2023deep}. Despite the fruitful results, most of the existing studies have not considered the motion uncertainty of HDVs, which is one of the key factors affecting the feasibility of the crossing order under mixed traffic. In this regard, game-theoretic scheduling methods have been proposed for exploration \cite{chandra2022gameplan}, \cite{pruekprasert2019decision}. However, it is still extremely challenging to properly model heterogeneous vehicle interactions with uncertainty, followed by efficiently searching for the optimal crossing order.

In terms of trajectory planning, the basic idea is to construct and solve optimal control problems (OCPs) with objectives such as time efficiency, comfort, and energy saving, to generate physically constrained and collision-free acceleration or velocity profiles for CAVs, while ensuring that the scheduled crossing order is followed. 
One approach is to formulate continuous OCPs and derive analytic solutions using Hamiltonian analysis to compute CAV trajectories in a sequential distributed fashion \cite{zhang2019decentralized}, \cite{malikopoulos2018decentralized}. However, in order to fulfill the control and state constraints, multiple cases have to be discussed sequentially for each CAV trajectory computation, which can be time-consuming; moreover, analytic solutions are not guaranteed to be derived when the constraints or cost functions are complex. A more common approach is to formulate discrete OCPs and solve them numerically using existing optimization methods; both centralized \cite{hult2018miqp}, \cite{murgovski2015convex}, \cite{karlsson2023optimal} and distributed \cite{de2017traffic}, \cite{bian2020cooperation}, \cite{pan2023hierarchical}. For example, Hult et al. introduced time variables to express the collision avoidance constraints as state couplings between different vehicles, planned CAV trajectories by solving a nonlinear program (NLP) \cite{hult2018miqp}, and proposed a semi-distributed solution to improve computational efficiency \cite{hult2019optimal}. In \cite{de2017traffic}, attraction sets were derived using reachability analysis tools to estimate potential collision times, and CAV trajectories were obtained by sequentially solving a set of quadratic programs (QPs). Zhang et al. \cite{zhang2024hierarchical} first centrally formulated non-overlapping spatio-temporal corridors as collision avoidance domains, and then computed trajectories for each CAV in a parallel distributed manner. In \cite{gong2024collision}, collision-free trajectories were trained offline based on iterative learning for different crossing modes. Distinguishing from the above studies oriented to fully CAV conditions, in \cite{jiang2023coordination}, CAVs in entry lanes were controlled to coordinate mixed platoons, with the assumption that HDVs are guided by advanced driver assistance systems. In \cite{faris2022optimization}, CAVs were coordinated under mixed traffic using model predictive control (MPC), and the infeasibility problem due to HDV uncertainty was addressed by relaxing a portion of the collision avoidance constraints.

Overall, the research on coordinated trajectory planning at unsignalized intersections is impressive, but the following limitations remain:
\begin{itemize}
\item[1)] To comprehensively address vehicle crossing, following, merging, and diverging conflicts at unsignalized intersections, time-dependent collision avoidance constraints are often complex and nonconvex, making the OCP difficult to solve \cite{hult2022semidistributed}. A common response is to simplify such constraints by pre-estimating the relevant time variables \cite{yao2023two}, \cite{zhao2021bilevel}; however, additional steps are required and both feasibility and optimality suffer.
\item[2)] The related work for mixed traffic is quite scarce, especially in terms of robust collision avoidance that takes into account the motion uncertainty of HDVs.
\item[3)] In response to the problem of different speed limits on straights and curves, most existing studies deal with the straight ahead and turning processes separately, an approach that lacks optimality.
\item[4)] Anchoring the temporal control horizon is tricky when the travel times of CAVs are optimization variables rather than specified values \cite{hadjigeorgiou2023real}, \cite{pan2023convex}. The vast majority of existing studies empirically set temporal control horizons for specific cases, which is clearly not suitable for practical applications.
\end{itemize}

\subsection{Contribution of the Paper}

In this paper, we propose a coordinated centralized MPC for trajectory planning of CAVs at an unsignalized intersection in mixed traffic. By sampling in distance and using an exact change of variables, the coordinated control problem is defined in the spatial domain rather than in the temporal domain as in most existing studies, thus establishing convex collision avoidance constraints without approximation and eliminating the horizon anchoring problem. A motion prediction model with uncertainty for HDVs is developed, and prediction deviations are compensated via MPC. This work is an extension of the previous study in \cite{murgovski2015convex}, which was limited to open-loop control for vehicle crossing scenarios under fully CAV conditions. The main contributions of this paper are summarized as follows:
\begin{itemize}
\item[1)] The coordinated control problem is formulated in the spatial domain as an NLP, which expresses the collision avoidance constraints for vehicle crossing, following, merging, and diverging in a unified linear form, and is able to handle spatially varying speed limits linearly.
\item[2)] The motion uncertainty of HDVs is decoupled and modeled as path uncertainty and speed uncertainty, whereby the robustness of collision avoidance is ensured in both spatial and temporal dimensions.
\item[3)] A real-time iteration (RTI) scheme is developed for efficient implementation of MPC, achieving a favorable trade-off between computational effort and optimality.
\end{itemize}

\subsection{Organization of the Paper}

The remainder of this paper is organized as follows. Section \ref{II} formulates the coordinated control problem at an unsignalized intersection under mixed traffic in the temporal domain. In Section \ref{III}, the problem formulation is transformed from the temporal domain to the spatial domain, the motion uncertainty of HDVs is modeled, robust collision avoidance constraints are established, and the coordinated control problem is reformulated as an NLP. In Section \ref{IV}, the NLP is deployed in the MPC framework, cost functions are presented, and an RTI scheme is developed for efficient computation. Section \ref{V} demonstrates the efficacy of the proposed methods through simulation case studies. Finally, Section \ref{VI} provides concluding remarks and future research directions.

\section{Problem Formulation}\label{II}
	
This paper focuses on the coordinated control of CAVs at an unsignalized intersection in mixed traffic, i.e., traffic with both CAVs and HDVs. In this section, we introduce the intersection scenario, model vehicle kinematics, and formulate the coordinated control problem.
	
\subsection{Unsignalized Intersection Scenario}
	
Mathematically, any unsignalized intersection can be represented by the tuple 
\begin{equation}
\left(p,x_l(p),y_l(p),\kappa_l(p),\psi_l(p),v_l^{\lim}(p),\mathcal L\right),
\end{equation}
where $p$ is a sampling variable denoting the distance traveled along the path, $x_l(p)$, $y_l(p)$ are the global coordinates of the predefined path $l\in\mathcal L$ at sample $p$, and $\kappa_l(p)$, $\psi_l(p)$, and $v_l^{\lim}(p)$ are the path curvature, tangent angle, and speed limit, respectively.
	
\begin{figure}[htbp]
\centering
\includegraphics[width=6cm,height=6.01cm]{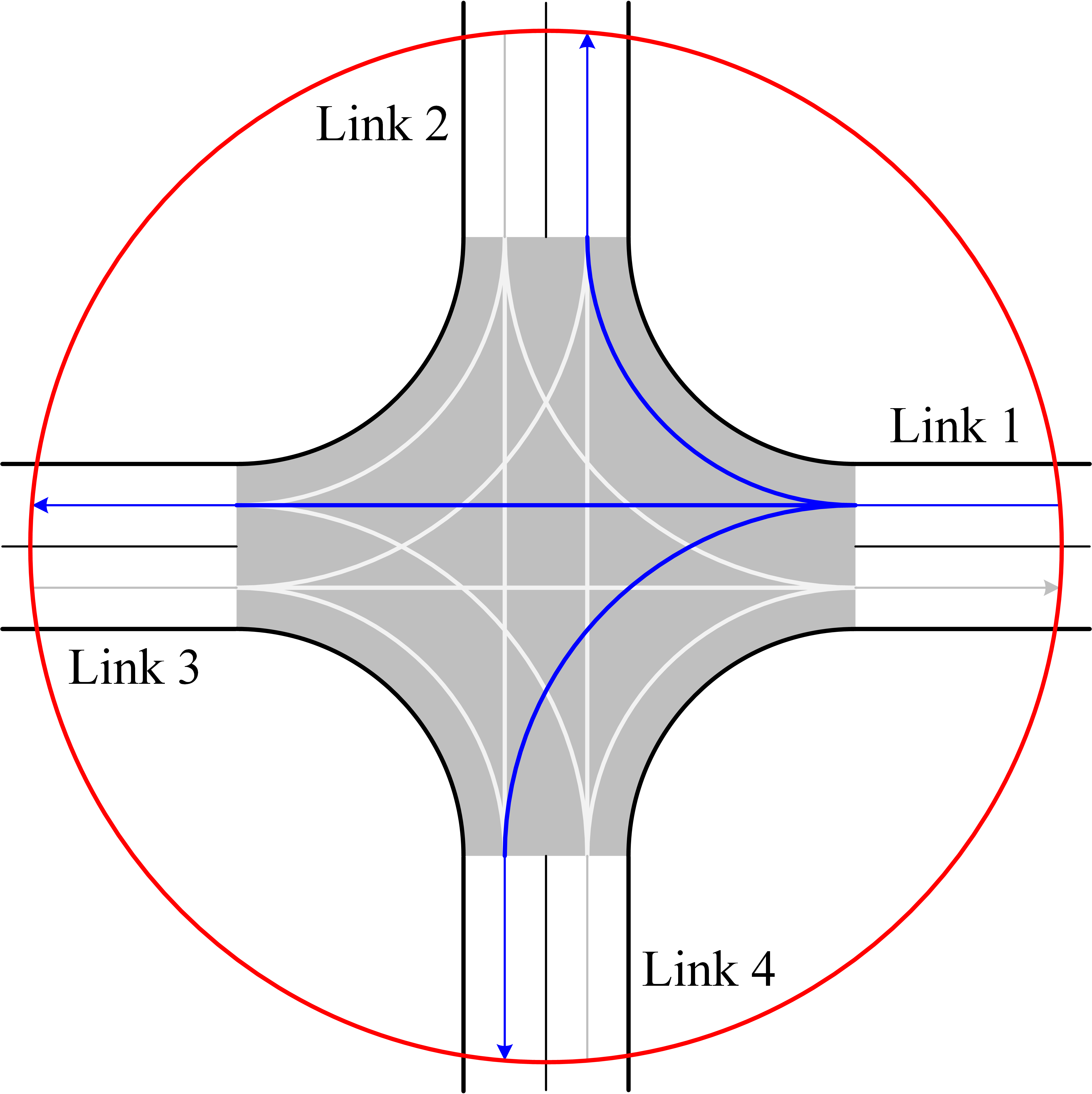}
\caption{An unsignalized intersection with four links, each with one entry lane and one exit lane. There are a total of 12 predefined paths, and the three paths starting from the first link are depicted in blue. The gray area is the central area. The red circle indicates the control boundary.}
\label{f1}
\end{figure}

As a typical example, consider the four-way unsignalized intersection shown in Fig. \ref{f1}, where vehicles are allowed to go straight, turn left, or turn right in the central area, depicted in gray. Each link consists of one entry lane and one exit lane, with the direction of travel shown by the arrows. Each entry lane leads to three paths to each exit lane that does not share a link with that entry lane, resulting in a total of 12 predefined paths. Finally, the red circle marks the control boundary, within which any vehicle is considered part of the intersection.
	
Now, consider $N\geq1$ vehicles, consisting of $M$ CAVs, with $1\leq M\leq N$, and $N-M$ HDVs, traveling through the intersection, and assume that 
\begin{itemize}
\item the crossing order has been given by a high-level scheduling module;
\item the states (location, speed, etc.) of all vehicles can be captured at any time;
\item the turning intentions of all vehicles are known and do not change;
\item all CAVs are under centralized control;
\item communication delays are negligible;
\item lane changing is prohibited. 
\end{itemize}
In practical terms, HDVs are sensed by CAVs and roadside units (RSUs), traffic information is shared through V2X communication, and centralized control is served by a certain CAV or RSU. 

The set of all vehicle indices is denoted as $\mathcal N=\{1,\dots,M,\dots,N\}$, and its subset $\mathcal M=\{1,\dots,M\}$ is defined as the set of CAV indices, and thus the set of HDV indices is $\mathcal {N\setminus M}$. Note that the intersection illustrated in Fig. \ref{f1} is only an example used herein, and the methods herein can be applied to unsignalized intersections with different geometries.
	
\subsection{Vehicle Paths}\label{II-B}

It is assumed that each vehicle $i\in\mathcal N$ is following a reference path 
\begin{equation}
\bm p_{i\mathrm r}(p)=\begin{bmatrix}
p & x_{i\mathrm r}(p) & y_{i\mathrm r}(p) & \kappa_{i\mathrm r}(p) & \psi_{i\mathrm r}(p) & v_{i\mathrm r}^{\lim}(p)
\end{bmatrix},
\end{equation}
which is one of the predefined paths, i.e., $\bm p_{i\mathrm r}(p)=\begin{bmatrix}
p & x_l(p) & y_l(p) & \kappa_l(p) & \psi_l(p) & v_l^{\lim}(p)
\end{bmatrix}$ for some $l\in\mathcal L$. Thereby, the path of vehicle $i$ can be expressed as
\begin{align}
\bm p_i(p)&=\begin{bmatrix}
p & x_i(p) & y_i(p) & \kappa_i(p) & \psi_i(p) & v_i^{\lim}(p)
\end{bmatrix}\nonumber\\
&=\bm p_{i\mathrm r}(\tilde p)+\bm w_i(\tilde p), \ \bm w_i(\tilde p)\in\mathscr W_i(\tilde p), \ \tilde p\in[\tilde p_{i0},\tilde p_{i\mathrm f}],
\end{align}
where $\tilde p$ is the projection of sample $p$ onto the reference path, $\bm w_i(\tilde p)$ is the path-following deviation of vehicle $i$ and its range $\mathscr W_i(\tilde p)$ is a complete bounded subset of $\mathbb R^6$, and $\tilde p_{i0}$ and $\tilde p_{i\mathrm f}$ are the projections of the initial position $p_{i0}$ and the final position $p_{i\mathrm f}$ of vehicle $i$ onto the reference path, respectively. Furthermore, CAVs are assumed to follow their respective reference paths perfectly, i.e., $\mathscr W_i(\tilde p)\equiv\{\bm0\}$ for all $i\in\mathcal M$. In contrast, the path-following deviations of HDVs are non-negligible and uncertain. 

\subsection{Vehicle Kinematics}

For each vehicle $i\in\mathcal N$, let $\bm x_i(t)=\begin{bmatrix}
p_i(t) & \dot p_i(t)
\end{bmatrix}^\top$ denote its state vector, where $t$ is the time, and $p_i(t)$ and $\dot p_i(t)$ are the longitudinal position and longitudinal velocity along its path, respectively. Then, vehicle $i$ is represented by a linear system
\begin{equation}
\dot{\bm x}_i(t)=\mathbf A\bm x_i(t)+\mathbf Bu_i(t),
\end{equation}
where
\begin{equation}\label{5}
\mathbf A=\left[
\begin{array}{cc}
0 & 1 \\ 0 & 0 
\end{array}
\right], \ 
\mathbf B=\left[
\begin{array}{cc}
0 \\ 1
\end{array}
\right],
\end{equation}
and the longitudinal acceleration along its path serves as the control signal, i.e., $u_i(t)=\ddot p_i(t)$.

The predefined path, combined with the longitudinal acceleration and longitudinal velocity profiles along it, completely defines the trajectory of a CAV. This trajectory can be tracked in real-time based on fine-grained vehicle dynamics to generate actionable control commands, specifically drive torque, braking force, and steering angle. Nevertheless, trajectory tracking is beyond the scope of this paper, which focuses on trajectory planning.

\subsection{State and Control Constraints}
	
Each CAV $i\in\mathcal M$ is subject to state and control constraints
\begin{equation}
\bm x_i(t)\in\left[\bm x_i^{\min},\bm x_i^{\max}(p_i(t))\right],
\end{equation}
\begin{equation}
u_i(t)\in\left[u_i^{\min},u_i^{\max}\right],
\end{equation}
where the inequalities are imposed for all $t\in[0,t_{i\mathrm f}]$ and
\begin{equation}
\begin{split}
&\bm x_i^{\min}=\begin{bmatrix}
p_{i0} & 0
\end{bmatrix}^\top,\\
&\bm x_i^{\max}(p_i(t))=\begin{bmatrix}
p_{i\mathrm f} & v_i^\mathrm{max}(p_i(t))
\end{bmatrix}^\top.
\end{split}
\end{equation}
The final time $t_{i\mathrm f}$, when CAV $i$ reaches its final position $p_{i\mathrm f}$, is free.
	
Unlike going straight, when a vehicle turns left or right, it moves on a path segment with significant curvature. In this case, the vehicle is subjected to non-negligible lateral forces, which should be limited to an acceptable level for driving safety and comfort. Since the lateral forces mainly provide centripetal acceleration, the maximum longitudinal velocity of CAV $i$ is given as
\begin{equation}\label{9}
v_i^{\max}(p_i(t))=\min\left\{v_i^{\lim}(p_i(t)),\sqrt{a_{i\mathrm c}^\mathrm{max}/\kappa_i(p_i(t))}\right\},
\end{equation}
where $a_{i\mathrm c}^{\max}$ is the maximum acceptable centripetal acceleration of CAV $i$. 
 
\subsection{Problem Statement}

Coordinated control aims to generate safe, efficient, and smooth trajectories for CAVs traveling through an unsignalized intersection in mixed traffic. For each vehicle $i\in\mathcal N$, let $\bm x_{i0}=\begin{bmatrix}
p_{i0} & v_{i0}
\end{bmatrix}^\top$ and $\bm x_{i\mathrm f}=\begin{bmatrix}
p_{i\mathrm f} & \text{free}
\end{bmatrix}^\top$ denote its initial and final states, respectively. Then, in the temporal domain, the coordinated control problem for CAVs is formulated as
\begin{subequations}\label{10}
\begin{align}
&\mathop{\min}_{\bm u_\mathrm C,\bm t_\mathrm{Cf}}\sum_{i=1}^{M}J_i(\bm x_i(t),u_i(t),\dot u_i(t),t_{i\mathrm f})\\
&\text{subject to}\nonumber\\ 
&\dot{\bm x}_i(t)=\mathbf A\bm x_i(t)+\mathbf Bu_i(t), \ \forall i\in\mathcal N,\label{10b}\\
&\bm x_i(t)\in\left[\bm x_i^{\min},\bm x_i^{\max}(p_i(t))\right], \ \forall i\in\mathcal M,\label{10c}\\
&u_i(t)\in\left[u_i^{\min},u_i^{\max}\right], \ \forall i\in\mathcal M,\label{10d}\\
&h(\bm x,\bm u_\mathrm C, \bm u_\mathrm H,\bm w)\leq0, \ \forall\bm u_\mathrm H\in\mathscr U_\mathrm H, \ \forall\bm w\in\mathscr W,\label{10e}\\
&\bm x_i(0)=\bm x_{i0}, \ \bm x_i(t_{i\mathrm f})=\bm x_{i\mathrm f}, \ \forall i\in\mathcal N,
\end{align}
\end{subequations}
where the constraints \eqref{10b}-\eqref{10d} are imposed for all $t\in[0,t_{i\mathrm f}]$. The optimization variables are the control signals and final times of all CAVs, i.e., $\bm u_\mathrm C=\begin{bmatrix}
u_1(t) & \dots & u_M(t)
\end{bmatrix}$ and $\bm t_\mathrm {Cf}=\begin{bmatrix}
t_{1\mathrm f} & \dots & t_{M\mathrm f}
\end{bmatrix}$, while those of all HDVs are uncertain variables. The robust constraint \eqref{10e} describes the collision avoidance conditions both between CAVs and between CAVs and HDVs, where $\bm x=\begin{bmatrix}
\bm x_1^\top(t) & \dots & \bm x_N^\top(t)
\end{bmatrix}$ and $\bm w=\begin{bmatrix}
\bm w_1(\tilde p_1(t)) & \dots & \bm w_N(\tilde p_N(t))
\end{bmatrix}$ represent the state vectors and path-following deviations of all vehicles, respectively, with $\bm w$ ranging over some bounded set $\mathscr W$, and $\bm u_\mathrm H$ represents the control signals of all HDVs, which belongs to some bounded set $\mathscr U_\mathrm H$. The cost function $J_i(\cdot)$ for each CAV $i\in\mathcal M$ may include penalties for deviation from the reference speed, for the control signal, for changes in the control signal, for the final time, for the final state, etc. The specific formulation of the cost function is deferred to Section \ref{IV-B}.

It can be seen that the collision avoidance constraint \eqref{10e} is the core of the optimization problem \eqref{10}. However, the explicit expression of the constraint \eqref{10e} in the temporal domain is often nonconvex and, due to the uncertainty involved in the motion of HDVs, complex in form. Furthermore, since the upper speed bound \eqref{9} is decision-variable dependent, the state constraints \eqref{10c} tend to be nonlinear and nonconvex. In addition, anchoring the temporal control horizon is tricky because the final times $\bm t_\mathrm {Cf}$ are optimization variables. These make problem \eqref{10} difficult to solve. Therefore, in the following section, we reformulate problem \eqref{10} as a standard NLP in the spatial domain.

\section{Modeling of Conflict Resolution Considering Motion Uncertainty}\label{III}

In this section, we transform the problem formulation from the temporal domain to the spatial domain using a non-approximate change of variables. Based on this, the motion uncertainty of HDVs is modeled, and robust collision avoidance constraints are further established. Finally, the coordinated control problem \eqref{10} is reformulated as an NLP.

\subsection{Temporal-Spatial Domain Transformation}

From a spatial perspective, the vehicle path is sampled in distance via the sampling variable $p$, while the travel time for the vehicle to reach any subsequent sample is unknown. In view of this, for each vehicle $i\in\mathcal N$, its travel time, $t_i(p)$, is chosen as a state and satisfies $t'_i(p) = 1/v_i(p)$. Here, the shorthand notation $(\cdot)'$ denotes the derivative with respect to distance, i.e., $(\cdot)'=\mathrm d(\cdot)/\mathrm dp$, and $v_i(p)$ is the speed of vehicle $i$ at sample $p$. Further, the variable representing the inverse vehicle speed, $z_i(p)=1/v_i(p)$, is introduced as another state, which is referred to as lethargy, as in \cite{murgovski2015convex}. Now, with $\bm{\tilde x}_i(p)=\begin{bmatrix} 
t_i(p) & z_i(p)
\end{bmatrix}^\top$ as the state vector, vehicle $i$ can be represented in the spatial domain by the linear system
\begin{equation}\label{11}
\bm{\tilde x}'_i(p)=\mathbf A\bm{\tilde x}_i(p)+\mathbf B\tilde u_i(p),
\end{equation}
where the spatial derivative of lethargy serves as the control signal, i.e., $\tilde u_i(p)=z'_i(p)$, and the matrices $\mathbf A$ and $\mathbf B$ remain as in \eqref{5}.

For each CAV $i\in\mathcal M$, the state constraint in the spatial domain is expressed as 
\begin{equation}
\bm{\tilde x}_i(p)\in\left[\bm{\tilde x}_i^{\min}(p),\bm{\tilde x}_i^{\max}\right),
\end{equation}
where the inequalities are imposed for all $p\in[p_{i0},p_{i\mathrm f}]$ and
\begin{equation}
\bm{\tilde x}_i^{\min}(p)=\begin{bmatrix}
0 & 1/v_i^{\max}(p)
\end{bmatrix}^\top, \ \bm{\tilde x}_i^{\max}=\begin{bmatrix}
\infty & \infty
\end{bmatrix}^\top.
\end{equation}
Here, with the same principle as in \eqref{9}, the maximum speed of CAV $i$ is given as 
\begin{equation}
v_i^{\max}(p)=\min\left\{v_i^{\lim}(p),\sqrt{a_{i\mathrm c}^{\max}/\kappa_i(p)}\right\}.
\end{equation}

To obtain the limits of the spatial control signal, let $a_i(p)\in\left[a_i^{\min},a_i^{\max}\right]$ denote the longitudinal acceleration of CAV $i$ at sample $p$, with $a_i^{\min}\leq0$ and $a_i^{\max}\geq0$. Observe that
\begin{equation}
\begin{split}
&\tilde u_i(p)=z'_i(p)=\left(\frac{1}{v_i(p)}\right)'=-\frac{v'_i(p)}{v_i^2(p)}=-v'_i(p)z_i^2(p),\\
&v'_i(p)=\frac{\mathrm d}{\mathrm d(t_i(p))}v_i(p)\cdot\frac{\mathrm d}{\mathrm dp}t_i(p)=a_i(p)z_i(p),
\end{split}
\end{equation}
which can be combined to derive $\tilde u_i(p)=-a_i(p)z_i^3(p)$. Therefore, the exact translation of the control constraint into the spatial domain is given as
\begin{equation}
\tilde u_i(p)\in z_i^3(p)\big[-a_i^{\max},-a_i^{\min}\big],
\end{equation}
where the inequalities are imposed for all $p\in[p_{i0},p_{i\mathrm f}]$.

Also, the initial and final states of each vehicle $i\in\mathcal N$ are translated into the spatial domain as $\bm{\tilde x}_{i0}=\begin{bmatrix}
0 & 1/v_{i0}
\end{bmatrix}^\top$ and $\bm{\tilde x}_{i\mathrm f}=\begin{bmatrix}
\text{free} & \text{free}
\end{bmatrix}^\top$, respectively.

\subsection{Motion Uncertainty Modeling}

\begin{figure}[htbp]
\centering
\includegraphics[width=7.5cm,height=5.84cm]{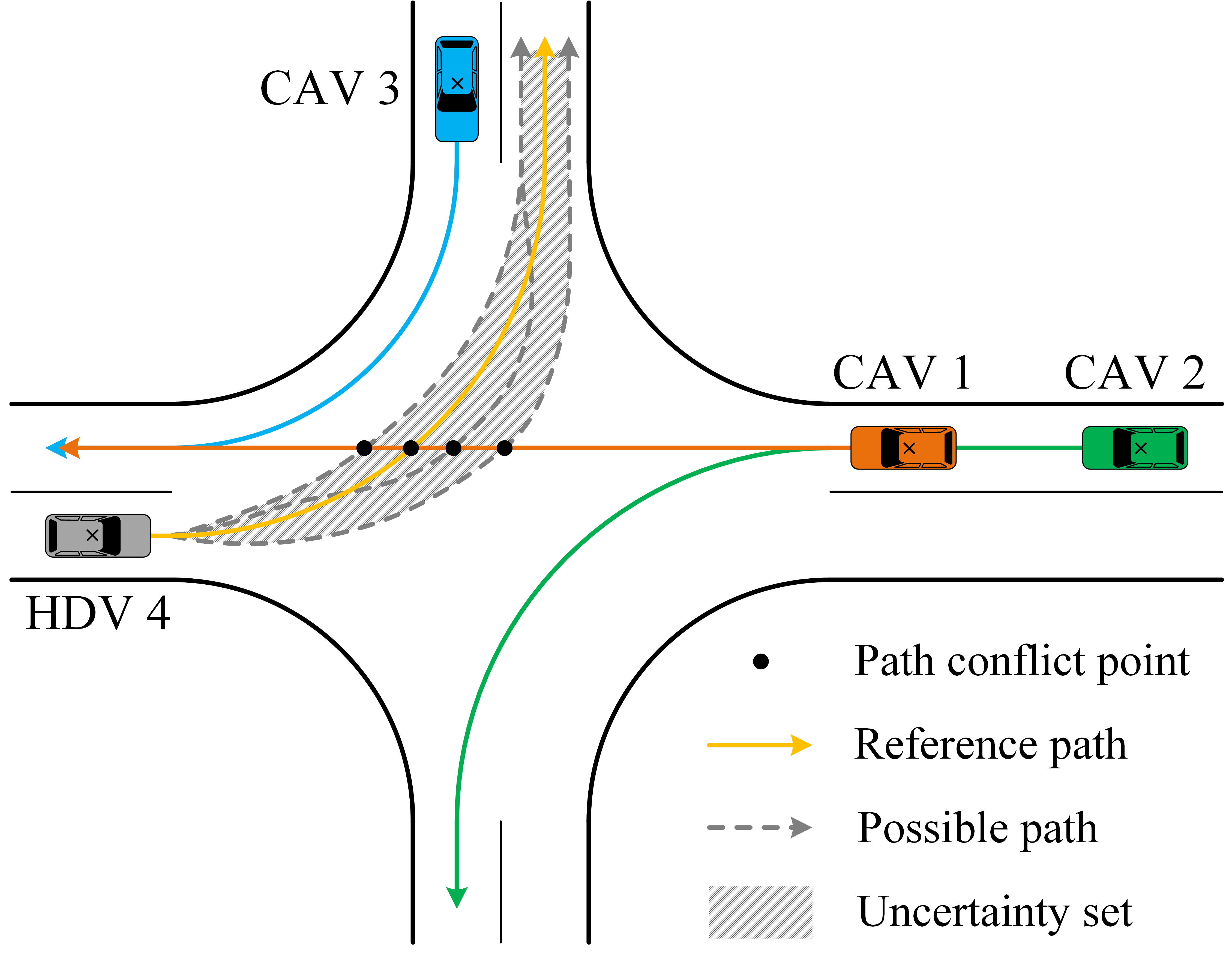}
\caption{Vehicles traveling through an unsignalized intersection. Each vehicle has a given reference path, depicted by a solid line with an end arrow. CAVs are assumed to follow their respective reference paths perfectly, while HDVs may not. The gray dashed lines indicate some of the possible paths of HDV 4, which are contained within an uncertainty set, depicted by the shaded area. The path conflict point between CAV 1 and HDV 4 is not fixed.}
\label{f2}
\end{figure}

The motion uncertainty of HDVs is reflected in both path and speed, which affect the location and timing of potential collisions. Taking the scenario depicted in Fig. \ref{f2} as an example, the left-turning HDV 4 may not strictly follow the reference path, but rather its possible paths are contained within an uncertainty set, as shown in the shaded area. Thus, the collision location between CAV 1 and HDV 4 is uncertain, as intuitively manifested by the non-fixed path conflict point. In contrast, since CAVs are assumed to follow their respective reference paths perfectly, the collision locations between them are predetermined.

\begin{figure}[htbp]
\centering
\includegraphics[width=7cm,height=5.23cm]{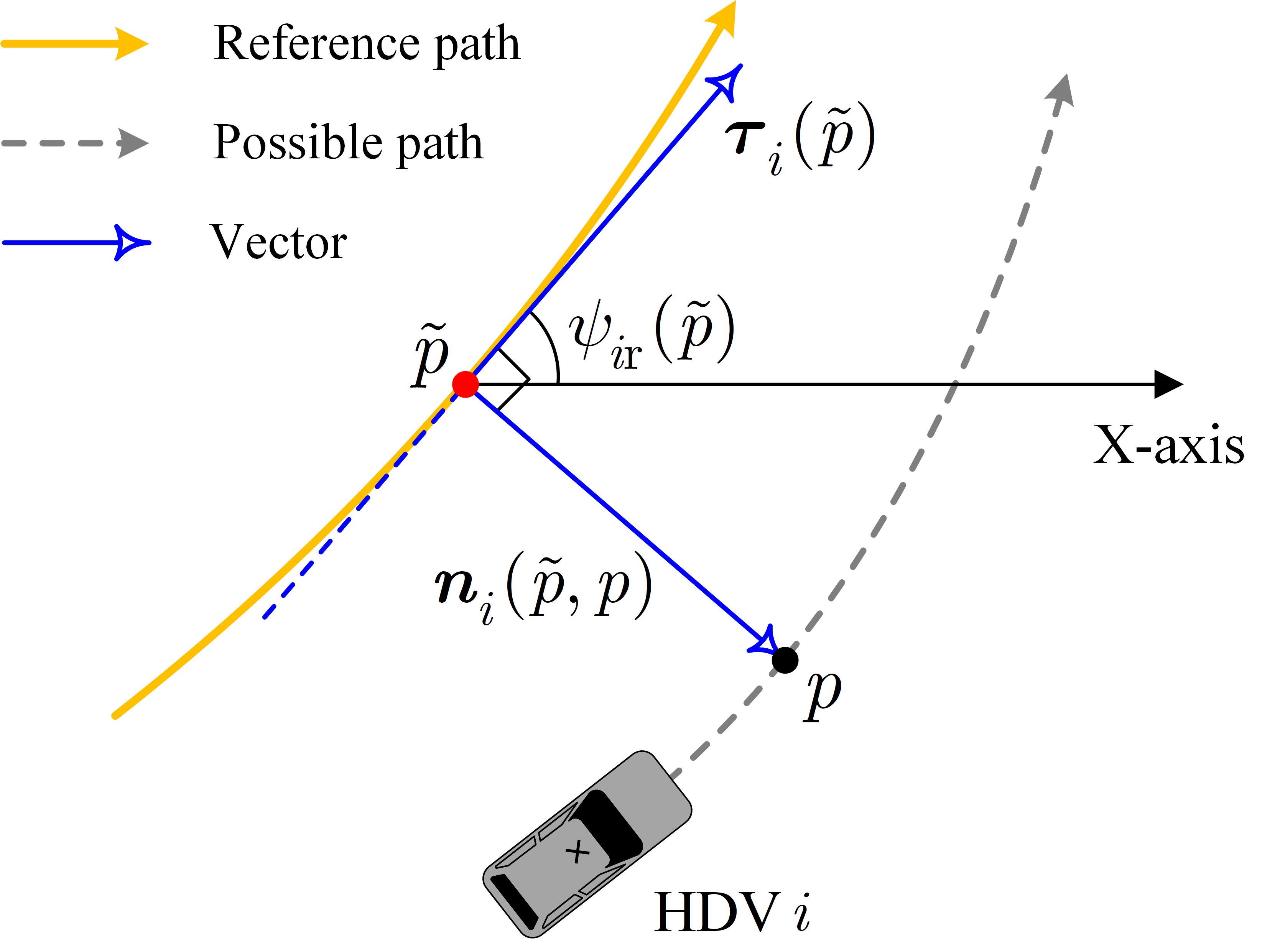}
\caption{Schematic of the path projection. The black dot indicates a sample and the red dot indicates the projection of that sample onto the reference path.}
\label{f3}
\end{figure}

As described in Section \ref{II-B}, the path of each vehicle $i\in\mathcal N$ can be expressed as $\bm p_i(p)=\bm p_{i\mathrm r}(\tilde p)+\bm w_i(\tilde p)$, where $\bm w_i(\tilde p)\in\mathscr W_i(\tilde p)$ and $\tilde p\in[\tilde p_{i0},\tilde p_{i\mathrm f}]$. According to projective geometry, the tangent vector at projection $\tilde p$ is denoted as $\bm\tau_i(\tilde p)=\begin{bmatrix} 
\cos\psi_{i\mathrm r}(\tilde p) & \sin\psi_{i\mathrm r}(\tilde p)
\end{bmatrix}$, and the vector from projection $\tilde p$ to sample $p$ is denoted as $\bm n_i(\tilde p,p)=\begin{bmatrix} 
x_i(p)-x_{i\mathrm r}(\tilde p) & y_i(p)-y_{i\mathrm r}(\tilde p)
\end{bmatrix}$, with $\bm\tau_i(\tilde p)\cdot\bm n_i(\tilde p,p)=0$, as shown in Fig. \ref{f3}. From this, given the initial coordinates $x_i(p_{i0})$, $y_i(p_{i0})$ of vehicle $i$, the lower bound $\tilde p_{i0}$ of the projection variable $\tilde p$ can be calculated, while the upper bound $\tilde p_{i\mathrm f}$ is obviously the length of the reference path. Further, let the offset, $\xi_i(\tilde p)$, of vehicle $i$ with respect to the reference path characterize its path uncertainty, defined as
\begin{equation}
\begin{split}
\xi_i(\tilde p)=\|\bm n_i(\tilde p,p)\|\cdot\operatorname{sgn}_\mathrm v(\bm\tau_i(\tilde p),\bm n_i(\tilde p,p)),\\
\xi_i(\tilde p)\in\mathscr U_i(\tilde p), \ \tilde p\in[\tilde p_{i0},\tilde p_{i\mathrm f}],
\end{split}
\end{equation}
where $\operatorname{sgn}_\mathrm v(\cdot)$ is a sign function with value domain $\{-1,1\}$, $\xi_i(\tilde p)<0$ means that the offset is clockwise, $\xi_i(\tilde p)>0$ means that the offset is counterclockwise, and the path uncertainty set $\mathscr U_i(\tilde p)$ is a complete bounded subset of $\mathbb R$. For each HDV $i\in\mathcal N\setminus\mathcal M$, $\mathscr U_i(\tilde p)$ is modeled based on its initial offset and road geometry, see Appendix \ref{A}; while for each CAV $i\in\mathcal M$, $\mathscr U_i(\tilde p)\equiv\{0\}$.

For ease of presentation, let the initial position of each HDV $i\in\mathcal N\setminus\mathcal M$ be $p_{i0}=\tilde p_{i0}$. Based on the kinematics \eqref{11}, the speed of HDV $i$ is estimated as
\begin{equation}\label{18}
\hat v_i(p)=\min\{\max\{\alpha,\bar v_i(p)\},\beta_i\hat v_i^{\max}(p)\},
\end{equation}
where $\alpha$ is a suitably small positive constant used as the lower limit of the speed estimate to avoid numerical problems due to excessively large values of lethargy, $\beta_i=\max\{1,v_{i0}/\hat v_i^{\max}(p_{i0})\}$ is the relaxation factor for the upper limit of the speed estimate, and
\begin{equation}\label{19}
\begin{split}
&\hat v_i^{\max}(p)=\min\left\{v_{i\mathrm r}^{\lim}(\bar p),\sqrt{\hat a_{i\mathrm c}^{\max}/\kappa_{i\mathrm r}(\bar p)}\right\},\\
&\bar p=\min\{p,\tilde p_{i\mathrm f}\},
\end{split}
\end{equation}
\begin{equation}
\bar v_i(p) =\sqrt{2\int_{p_{i0}}^{p}\gamma_i(s)\mathrm ds+v_{i0}^2}, \ \gamma_i(p)\in\Gamma_i(p).
\end{equation}
Here, $s$ is the integral variable corresponding to the sampling variable $p$, and $\gamma_i(p)$ is the perturbed acceleration of HDV $i$, with its range $\Gamma_i(p)$ being a complete bounded subset of $\mathbb R$. The upper limit of the speed estimate, $\hat v_i^{\max}(p)$, is derived by approximating the actual path speed limit and curvature using those of the reference path, as in \eqref{19}, and is corrected in \eqref{18} by the relaxation factor $\beta_i$.

Further, the travel time for HDV $i$ to reach the normal at projection $\tilde p$ is estimated as
\begin{equation}\label{21}
\begin{split}
\hat t_i(\tilde p)=\int_{p_{i0}}^{\tilde p+\omega_i(\tilde p)}\frac{1}{\hat v_i(p)}\mathrm dp, \ \omega_i(\tilde p)\in\Omega_i(\tilde p),\\
\hat t_i(\tilde p)\in\left[\hat t_i^{\min}(\tilde p),\hat t_i^{\max}(\tilde p)\right], \ \tilde p\in[\tilde p_{i0},\tilde p_{i\mathrm f}],
\end{split}
\end{equation}
where $\omega_i(\tilde p)$ is the travel distance deviation of HDV $i$ with respect to the reference path, i.e., $p=\tilde p+\omega_i(\tilde p)$, and its range $\Omega_i(\tilde p)$ is a complete bounded subset of $\mathbb R$. From \eqref{18}-\eqref{21}, it can be seen that the uncertainty in the travel time of HDV $i$ is jointly characterized by the travel distance deviation $\omega_i(\tilde p)$ and the perturbed acceleration $\gamma_i(p)$. Moreover, the minimum travel time estimate, $\hat t_i^{\min}(\tilde p)$, is obtained by making $\omega_i(\tilde p)=\min\Omega_i(\tilde p)$ and $\gamma_i(p)=\max\Gamma_i(p)$, and the maximum travel time estimate, $\hat t_i^{\max}(\tilde p)$, is obtained by making $\omega_i(\tilde p)=\max\Omega_i(\tilde p)$ and $\gamma_i(p)=\min\Gamma_i(p)$.

\subsection{Robust Collision Avoidance Constraints}

Vehicle collisions can occur where paths intersect, overlap, or approach; such areas are called critical zones. The collision avoidance constraints can be imposed by constraining the time when CAVs enter and exit the critical zone so that it is occupied by at most one vehicle at a time.

Let $\mathfrak O\in\mathcal N^N$ represent a given vehicle crossing order, which is a specific permutation of all elements in $\mathcal N$, and let $\mathfrak O_i$ denote the order of vehicle $i$ in $\mathfrak O$. Considering the physical dimensions and safety margins of the vehicle, let $\mathcal B_i(\tilde p,\xi_i(\tilde p))$ denote the rectangle bounding box of vehicle $i$ with projection $\tilde p$ and offset $\xi_i(\tilde p)$, and the heading angle of vehicle $i$ is approximated by the tangent angle $\psi_{i\mathrm r}(\tilde p)$ of the reference path. Then, for vehicle $i$ and vehicle $j$ with $\mathfrak O_i<\mathfrak O_j$, the possible collision locations are given as
\begin{align}
\mathcal P_{ij}=\big\{&(\tilde p_1,\tilde p_2)\in[\tilde p_{i0},\tilde p_{i\mathrm f}]\times[\tilde p_{j0},\tilde p_{j\mathrm f}] \mid\mathcal B_i(\tilde p_1,\xi_i(\tilde p_1))\nonumber\\
&\cap\mathcal B_j(\tilde p_2,\xi_j(\tilde p_2))\neq\emptyset, \ \xi_i(\tilde p_1)\in\mathscr U_i(\tilde p_1),\nonumber\\
&\xi_j(\tilde p_2)\in\mathscr U_j(\tilde p_2), \ \mathfrak O_i<\mathfrak O_j, \ \eqref{23}\big\},
\end{align}
\begin{equation}\label{23}
\nexists \mathfrak O_i<\mathfrak O_m<\mathfrak O_j,\ \bm p_{m\mathrm r}(p)=\bm p_{i\mathrm r}(p)\lor \bm p_{m\mathrm r}(p)=\bm p_{j\mathrm r}(p),
\end{equation}
where $\mathcal B_i(\cdot)\cap\mathcal B_j(\cdot)\neq\emptyset$ means that the bounding box of vehicle $i$ overlaps with that of vehicle $j$, which is regarded as a collision. The overlap determination is achieved using the separating axis theorem. The condition \eqref{23} is used to ensure the compactness of potential collision identification. For example, in vehicle following, it is not necessary to consider collision avoidance with vehicles other than the immediate leader.

To identify the critical zones for vehicle $i$ and vehicle $j$, suppose that the collision location $\tilde p_2$ of vehicle $j$ is the entrance of some critical zone. Then the exit is given by
\begin{equation}
f(\tilde p_2)=\max\{\tilde p_1\mid(\tilde p_1,\tilde p_2)\in\mathcal P_{ij}\}.
\end{equation}
Suppose that the collision location $\tilde p_1$ of vehicle $i$ is the exit of some critical zone. Then the entrance is given by 
\begin{equation}
g(\tilde p_1)=\min\{\tilde p_2\mid(\tilde p_1,\tilde p_2)\in\mathcal P_{ij}\}.
\end{equation}
Thus, the critical zones can be represented by their exit and entrance locations
\begin{equation}
\mathcal C_{ij}=\{(\tilde p_\mathrm{out},\tilde p_\mathrm {in})\in\mathcal P_{ij}\mid\tilde p_\mathrm{out}=f(\tilde p_\mathrm{in}), \ \tilde p_\mathrm{in}=g(\tilde p_\mathrm{out})\}.
\end{equation}

\begin{figure}[htbp]
\centering	\includegraphics[width=7.5cm,height=6.06cm]{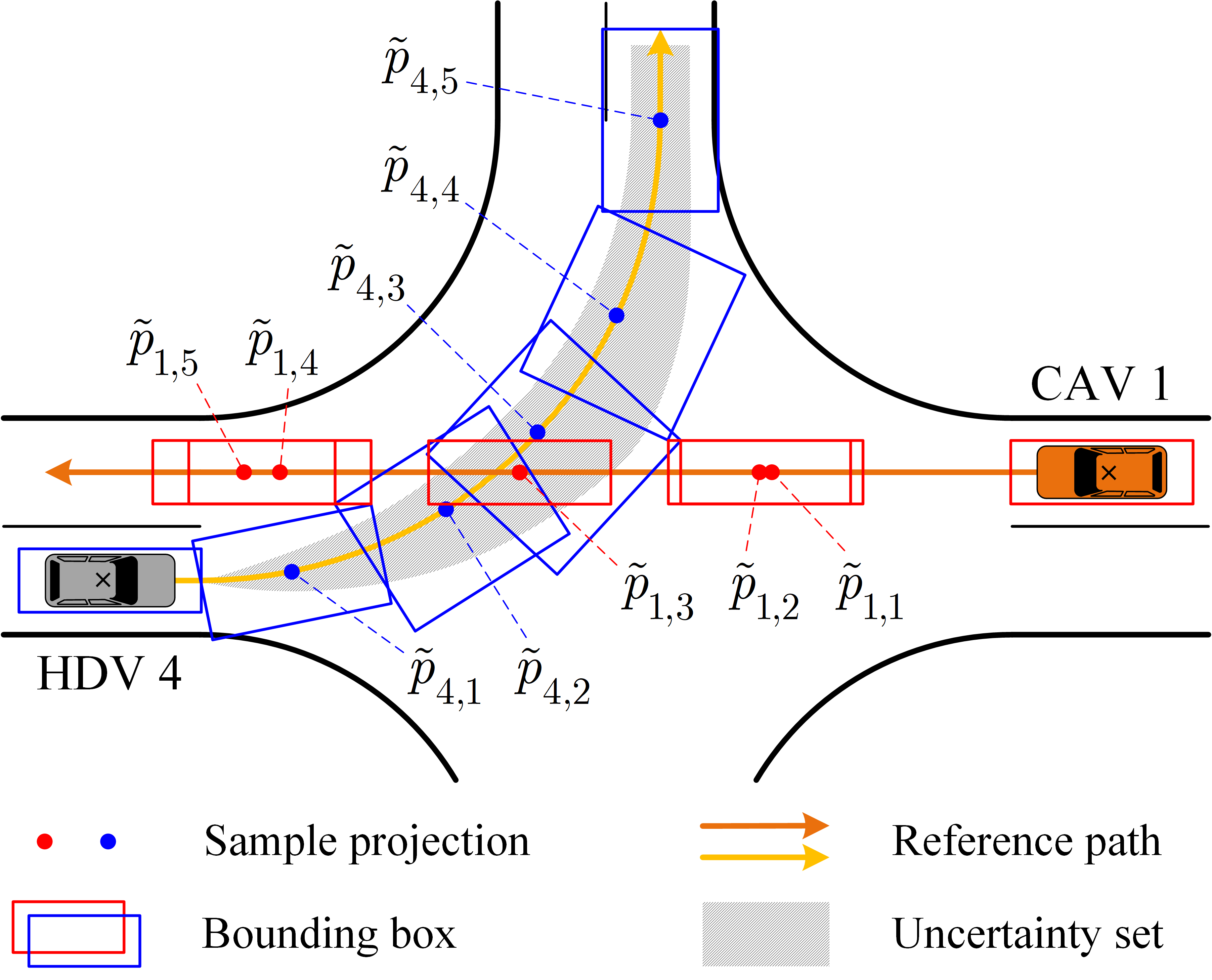}
\caption{Illustration of the traffic conflict between CAV 1 and HDV 4 in the scenario shown in Fig. \ref{f2}. The path uncertainty of HDV 4 is reflected in its bounding box, which becomes wider as the path uncertainty increases.}
\label{f4}
\end{figure}

Now, the collision avoidance constraints can be expressed as linear constraints
\begin{equation}\label{27}
\begin{split}
t_{ij}(\tilde p_\mathrm{out},\tilde p_\mathrm{in})+\lambda_{ij}(\tilde p_\mathrm{in})\leq-t_j^\mathrm{des}, \ \lambda_{ij}(\tilde p_\mathrm{in})\in\left[-t_j^\mathrm{des},0\right],\\
\forall(\tilde p_\mathrm{out},\tilde p_\mathrm{in})\in\mathcal C_{ij}, \ \forall i,j\in\mathcal N, \ \{i,j\}\nsubseteq\mathcal N\setminus\mathcal M,
\end{split}
\end{equation}
where $\lambda_{ij}(\cdot)$ is the slack variable introduced to avoid infeasibility problems caused by motion prediction deviations for HDVs, $t_j^\mathrm{des}$ is the positive, desired time gap of vehicle $j$, and
\begin{align}\label{28}
t_{ij}&(\tilde p_\mathrm{out},\tilde p_\mathrm{in})=\nonumber\\
&\begin{cases}
\hat t_i^{\max}(\tilde p_\mathrm{out})-t_j(\tilde p_\mathrm{in}), &\text{if} \ i\in\mathcal N\setminus\mathcal M, \ j\in\mathcal M,\\
t_i(\tilde p_\mathrm{out})-\hat t_j^{\min}(\tilde p_\mathrm{in}), &\text{if} \ i\in\mathcal M, \ j\in\mathcal N\setminus\mathcal M,\\
t_i(\tilde p_\mathrm{out})-t_j(\tilde p_\mathrm{in}), &\text{if} \ i,j\in\mathcal M.
\end{cases}
\end{align}
These constraints state that vehicle $i$ must exit the critical zone before vehicle $j$ enters it, for which the slack variable $\lambda_{ij}(\cdot)$ receives a substantial penalty in the objective function. Moreover, the constraints are robust because they cover the edge cases for both collision location and timing. As an example, Fig. \ref{f4} depicts in detail the traffic conflict between CAV 1 and HDV 4 in the scenario shown in Fig. \ref{f2}. It can be seen that the path uncertainty of HDV 4 is reflected in its bounding box, which becomes wider as the path uncertainty increases. If CAV 1 is specified to pass after HDV 4, i.e., $\mathfrak O_4<\mathfrak O_1$, then $(\tilde p_{4,3},\tilde p_{1,1})$ and $(\tilde p_{4,4},\tilde p_{1,2})$ are the only two critical zones plotted in Fig. \ref{f4}, which cover all the other plotted collision locations such as $(\tilde p_{4,1},\tilde p_{1,4})$ and $(\tilde p_{4,2},\tilde p_{1,3})$. Similarly, if CAV 1 is specified to pass before HDV 4, i.e., $\mathfrak O_1<\mathfrak O_4$, then $(\tilde p_{1,4},\tilde p_{4,1})$ and $(\tilde p_{1,5},\tilde p_{4,2})$ are the only two critical zones plotted in Fig. \ref{f4}. This indicates that the identified critical zones are full-coverage, ensuring the robustness of collision avoidance in the spatial dimension, while being non-redundant. For the timing of HDVs entering and exiting the critical zone, the worst-case scenario is always considered to ensure the robustness of collision avoidance in the temporal dimension. As in \eqref{28}, if an HDV is specified to go first, its maximum travel time estimate is adopted; otherwise, its minimum travel time estimate is adopted.

It should be noted that the collision avoidance constraints \eqref{27} are able to handle crossing, following, merging, and diverging conflicts involving CAVs in a unified linear form; however, conflicts between HDVs are not included as they cannot be directly resolved through the coordinated control of CAVs. This exclusion is handled by the notation $\{i,j\}\nsubseteq\mathcal N\setminus\mathcal M$ in \eqref{27}.

\subsection{Problem Reformulation}

In the spatial domain, for a given vehicle crossing order $\mathfrak O\in\mathcal N^N$, the coordinated control problem for CAVs is now reformulated as
\begin{subequations}\label{29}
\begin{align}
&\mathop{\min}_{\bm{\tilde u}_\mathrm C,\bm\lambda}J(\bm\lambda)+\sum_{i=1}^{M}\tilde J_i(\bm{\tilde x}_i(p),\tilde u_i(p),\tilde u'_i(p))\label{29a}\\
&\text{subject to}\nonumber\\ 
&\bm{\tilde x}'_i(p)=\mathbf A\bm{\tilde x}_i(p)+\mathbf B\tilde u_i(p), \ \forall i\in\mathcal M,\label{29b}\\
&\bm{\tilde x}_i(p)\in\left[\bm{\tilde x}_i^{\min}(p),\bm{\tilde x}_i^{\max}\right), \ \forall i\in\mathcal M,\label{29c}\\
&\tilde u_i(p)\in z_i^3(p)\big[-a_i^{\max},-a_i^{\min}\big], \ \forall i\in\mathcal M,\label{29d}\\
\begin{split}	
t_{ij}(\tilde p_\mathrm{out},\tilde p_\mathrm{in})+\lambda_{ij}(\tilde p_\mathrm{in})\leq-t_j^\mathrm{des}, \ \lambda_{ij}(\tilde p_\mathrm{in})\in\left[-t_j^\mathrm{des},0\right],\\
\forall(\tilde p_\mathrm{out},\tilde p_\mathrm{in})\in\mathcal C_{ij}, \ \forall i,j\in\mathcal N, \ \{i,j\}\nsubseteq\mathcal N\setminus\mathcal M,
\end{split}\\
&\bm{\tilde x}_i(p_{i0})=\bm{\tilde x}_{i0}, \ \forall i\in\mathcal M,
\end{align}
\end{subequations}
where the constraints \eqref{29b}-\eqref{29d} are imposed for all $p\in[p_{i0},p_{i\mathrm f}]$, and $\bm\lambda$ is a vector consisting of all the slack variables, which together with the control signals of all CAVs, $\bm{\tilde u}_\mathrm C=\begin{bmatrix}
\tilde u_1(p) & \dots & \tilde u_M(p)
\end{bmatrix}$, serve as the optimization variables. Assuming that the objective function in \eqref{29a} is nonlinear, problem \eqref{29} is a nonconvex NLP, and the nonconvexity stems from the control constraints \eqref{29d}, except for which all other constraints are linear. The NLP \eqref{29} can be solved iteratively using convex subproblems constructed via convex approximation, a solution method commonly referred to as sequential convex programming (SCP), or sequential quadratic programming (SQP) if the subproblems are QPs (whether convex or nonconvex). To ensure applicability in dynamic mixed traffic with uncertainty, program \eqref{29} is deployed in the MPC framework; see the following section.

\section{Centralized Model Predictive Control and Computationally Efficient Solutions}\label{IV}

In this section, the coordinated control problem \eqref{29} is discretized and written in a receding horizon fashion, convex quadratic cost functions are formulated, and an RTI scheme is developed using the inner approximation of the search space to improve computational efficiency.

\subsection{Receding Horizon Optimization}

In order to compensate for motion prediction deviations for HDVs and other possible disturbances, it is necessary to iteratively update and resolve the coordinated control problem. For each CAV $i\in\mathcal M$, let $\mathcal P_i=\{p_{i0},p_{i0}+\Delta p,\dots,p_{i\mathrm f}+\epsilon_i\}$ denote the set of its discrete distance samples corresponding to the continuous sample interval $[p_{i0},p_{i\mathrm f}]$, where $\Delta p$ is the distance sampling interval and $\epsilon_i\in[0,\Delta p)$ is a relaxation term introduced to guarantee uniform sampling. Let $k$ be the discrete sampling variable. Then, the coordinated control problem \eqref{29} is discretized as
\begin{subequations}\label{30}
\begin{align}
&\mathop{\min}_{\bm{\tilde u}_{\mathrm C}^{\mathrm d},\bm\lambda^\mathrm d}J^\mathrm d(\bm\lambda^\mathrm d)+\sum_{i=1}^{M}\tilde J_i^\mathrm d(\bm{\tilde x}_i(k),\tilde u_i(k))\\
&\text{subject to}\nonumber\\ 
&\bm{\tilde x}_i(k+\Delta p)=\mathbf A^\mathrm d\bm{\tilde x}_i(k)+\mathbf B^\mathrm d\tilde u_i(k), \ \forall i\in\mathcal M,\label{30b}\\
&\bm{\tilde x}_i(k)\in\left[\bm{\tilde x}_i^{\min}(k),\bm{\tilde x}_i^{\max}\right), \ \forall i\in\mathcal M,\label{30c}\\
&\tilde u_i(k)\in z_i^3(k)\big[-a_i^{\max},-a_i^{\min}\big], \ \forall i\in\mathcal M,\label{30d}\\
\begin{split}	
t_{ij}\big(\tilde k_\mathrm{out},\tilde k_\mathrm{in}\big)+\lambda_{ij}(\tilde k_\mathrm{in})\leq-t_j^\mathrm{des}, \ \lambda_{ij}(\tilde k_\mathrm{in})\in\left[-t_j^\mathrm{des},0\right],\\
\forall(\tilde k_\mathrm{out},\tilde k_\mathrm{in})\in\mathcal C_{ij}^\mathrm d, \ \forall i,j\in\mathcal N, \ \{i,j\}\nsubseteq\mathcal N\setminus\mathcal M,
\end{split}\\
&\bm{\tilde x}_i(p_{i0})=\bm{\tilde x}_{i0}, \ \forall i\in\mathcal M,
\end{align}
\end{subequations}
where the constraints \eqref{30b} and \eqref{30d} hold for all $k\in\mathcal P_i^-=\mathcal P_i\setminus\{p_{i\mathrm f}+\epsilon_i\}$, the constraints \eqref{30c} hold for all $k\in\mathcal P_i$, the superscript $(\cdot)^\mathrm d$ denotes the discretized version, and 
\begin{equation}
\mathbf A^\mathrm d=\left[
\begin{array}{cc}
1 & \Delta p \\ 0 & 1 
\end{array}
\right], \ 
\mathbf B^\mathrm d=\left[
\begin{array}{cc}
\dfrac{\Delta p^2}{2} & \Delta p
\end{array}
\right]^\top.
\end{equation}

Assume that all CAVs have synchronized clocks. After solving the discretized problem \eqref{30} each time, each CAV $i\in\mathcal M$ implements the first segment $\{\tilde u_i^*(p_{i0}),\dots,\tilde u_i^*(p_{i0}+(n_i-1)\Delta p)\}$ of its optimal control sequence over a pre-specified time sampling period $\Delta t$. Here, $n_i$ is the number of control steps for CAV $i$, defined as $n_i=\max\{n\in\{1,\dots,|\mathcal P_i^-|\}\mid t_i^*(p_{i0}+(n-1)\Delta p)<\Delta t\}$. Then, problem \eqref{30} is updated based on the new traffic state and crossing order (if any), and is re-solved over the shifted horizon.

Note that the proposed centralized MPC is iterated with the time sampling period $\Delta t$ but solved in the spatial domain. Consequently, the number of control steps may differ for the same CAV across iterations and for different CAVs in the same iteration. In addition, the control horizon for each CAV shrinks with increasing MPC iterations, indicating that the computational effort of the MPC algorithm decreases as CAVs approach their final locations and increases as new CAVs enter the intersection.

\subsection{Convex Quadratic Cost Functions}\label{IV-B}

Speed tracking and travel time reduction are commonly used cost functions for the temporal-domain problem \eqref{10}, and their convex quadratic formulations for the spatial-domain problem \eqref{30} are provided here.

\subsubsection{Speed Tracking}

The reference speed is tracked using the cost function 
\begin{equation}
\tilde J_i^\mathrm d(\cdot)=\tilde J_{i1}^\mathrm d(\bm{\tilde x}_i(k),\tilde u_i(k))+\tilde J_{i2}^\mathrm d(\bm{\tilde x}_i(k),\tilde u_i(k)),
\end{equation}
where $\tilde J_{i1}^\mathrm d(\cdot)$ and $\tilde J_{i2}^\mathrm d(\cdot)$ represent the costs within and beyond the actual finite prediction horizon, respectively, thereby effectively mimicking an infinite horizon. Specifically,
\begin{align}
\tilde J_{i1}^\mathrm d(\cdot)&=q_{i1}(z_i(p_{i\mathrm f}+\epsilon_i)-1/v_{i\mathrm r}(p_{i\mathrm f}+\epsilon_i))^2\nonumber\\
&+\sum_{k\in\mathcal P_i^-}\big[q_{i1}(z_i(k)-1/v_{i\mathrm r}(k))^2+r_i\tilde u_i^2(k)\nonumber\\
&+e_i(\tilde u_i(k)-\tilde u_i(k-\Delta p))^2\big],
\end{align}
where $v_{i\mathrm r}(\cdot)$ is the reference speed and $q_{i1}$, $r_i$, and $e_i$ are positive weights. Penalties for the control signal and its difference are included to mitigate sudden shifts in acceleration and jerk, thereby reducing discomfort and energy consumption. The cost beyond the actual finite prediction horizon (extending to infinity) is formulated as
\begin{equation}
\tilde J_{i2}^\mathrm d(\cdot)=\sum_{k\in \mathcal P_i^+}\left[q_{i1}(z_i(k)-1/v_{i\mathrm{r,s}})^2+r_i\tilde u_i^2(k)\right],
\end{equation}
where $\mathcal P_i^+=\{p_{i\mathrm f}+\epsilon_i,p_{i\mathrm f}+\epsilon_i+\Delta p,\dots,\infty\}$ and $v_{i\mathrm{r,s}}$ is the steady state reference speed. This cost is introduced to stabilize the second state of the system, $z_i(k)$. In the standard linear-quadratic regulator fashion, it can be shown that 
\begin{equation}
\tilde J_{i2}^\mathrm d(\cdot)=P_i(z_i(p_{i\mathrm f})-1/v_{i\mathrm {r,s}})^2,
\end{equation}
where $P_i$ is the solution to the discrete algebraic Riccati equation, explicitly expressed as
\begin{equation}
P_i=\frac{q_{i1}}{2}+\sqrt{\left(\frac{q_{i1}}{2}\right)^2+\frac{q_{i1}r_i}{\Delta p^2}}.
\end{equation}
\subsubsection{Travel Time Reduction}

The final travel time is minimized using the cost function
\begin{align}
\tilde J_i^\mathrm d(\cdot)&=q_{i2}t_i(p_{i\mathrm f}+\epsilon_i)\nonumber\\	
&+\sum_{k\in\mathcal P_i^-}\left[r_i\tilde u_i^2(k)+e_i(\tilde u_i(k)-\tilde u_i(k-\Delta p))^2\right],
\end{align}
where $q_{i2}$ is a positive weight and the summation term is a penalty for discomfort and energy consumption. At the traffic control level, this cost is often used to maximize throughput.

\subsubsection{Penalties for Slack Variables}
The cost of exploiting the slack variables is formulated as
\begin{equation}
J^\mathrm d(\bm\lambda^\mathrm d)=N_\mathrm sq_\mathrm s\|\bm\lambda^\mathrm d\|^2,
\end{equation}
where $N_\mathrm s$ is the number of the discretized slack variables and $q_\mathrm s$ is a large positive weight.

\subsection{Real-Time Iteration}

Although the NLP \eqref{30} can be solved with off-the-shelf solvers, the process might be time-consuming if the solver has to run until convergence. A computationally efficient solution is to stop the SCP/SQP scheme after only one iteration in each MPC update, which is known as RTI. In order to apply RTI, we need to ensure that the feasible solution to the first subproblem of the SCP/SQP scheme always yields a feasible solution to the original problem (even when the slack variables are not used).

Recalling the continuous control constraints \eqref{29d}, it is found that the lower bound is convex and the upper bound is concave. Therefore, the linearization of the control constraints \eqref{29d} is an inner approximation to the original search space, which means that feasible points within the linearized bounds are also feasible in the original nonconvex set. Using a first-order Taylor expansion, the constraints \eqref{29d} are linearized around the lethargy $z_{i,\mathrm{lin}}(p)$, which is then discretized to yield
\begin{equation}\label{38}
\tilde u_i(k)\in\left[\tilde u_i^{\min}(k,z_i(k)),\tilde u_i^{\max}(k,z_i(k))\right], \ \forall i\in\mathcal M,
\end{equation}
where
\begin{equation}
\begin{split}
&\tilde u_i^{\min}(k,z_i(k))=a_i^{\max}(2z_{i,\mathrm{lin}}(k)-3z_i(k))z_{i,\mathrm{lin}}^2(k),\\
&\tilde u_i^{\max}(k,z_i(k))=a_i^{\min}(2z_{i,\mathrm{lin}}(k)-3z_i(k))z_{i,\mathrm{lin}}^2(k).
\end{split}
\end{equation}
Now, by replacing the nonconvex constraints \eqref{30d} in the original problem \eqref{30} with the linear constraints \eqref{38}, the target subproblem is obtained, which is a computationally efficient convex QP. This allows the application of RTI. 

It should be noted that in the first MPC iteration for each CAV $i\in\mathcal M$, since there is no previous solution available, the inverse of the reference and maximum speeds are chosen to linearize the control constraint for the speed tracking cost and the travel time reduction cost, respectively. Thereafter, for both cost functions, the linearization is performed about the previous solution, shifted by $n_i$ samples.

\section{Simulation Results}\label{V}

\begin{figure}[htbp]
\centering
\subfloat[Scenario 1: Four vehicles consising of 3 CAVs and 1 HDV traveling through an unsignalized intersection.]{\includegraphics[width=4.2cm,height=4.2cm]{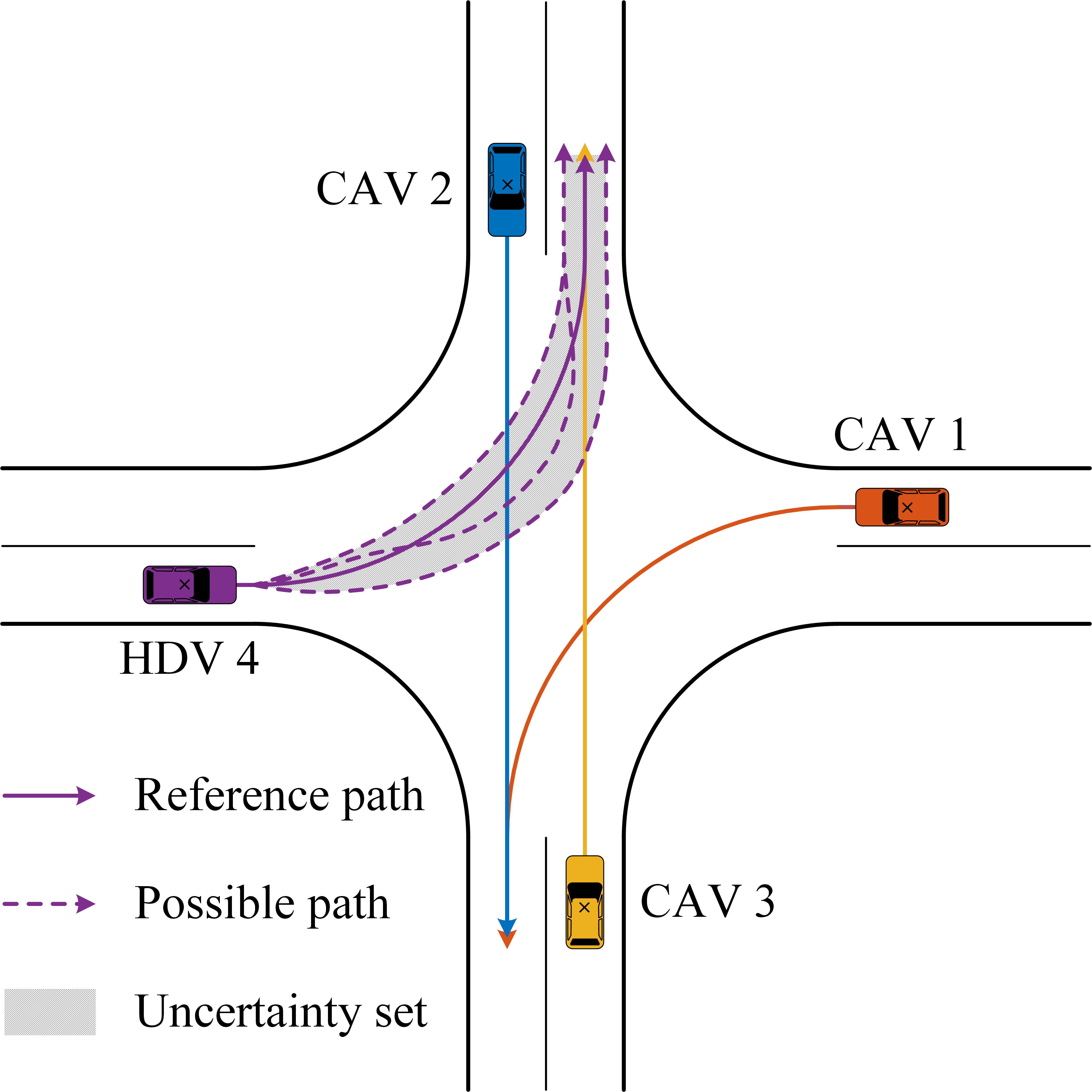}}
\hfill
\subfloat[Scenario 2: Eight vehicles consising of 6 CAVs and 2 HDVs traveling through an unsignalized intersection.]{\includegraphics[width=4.41cm,height=4.2cm]{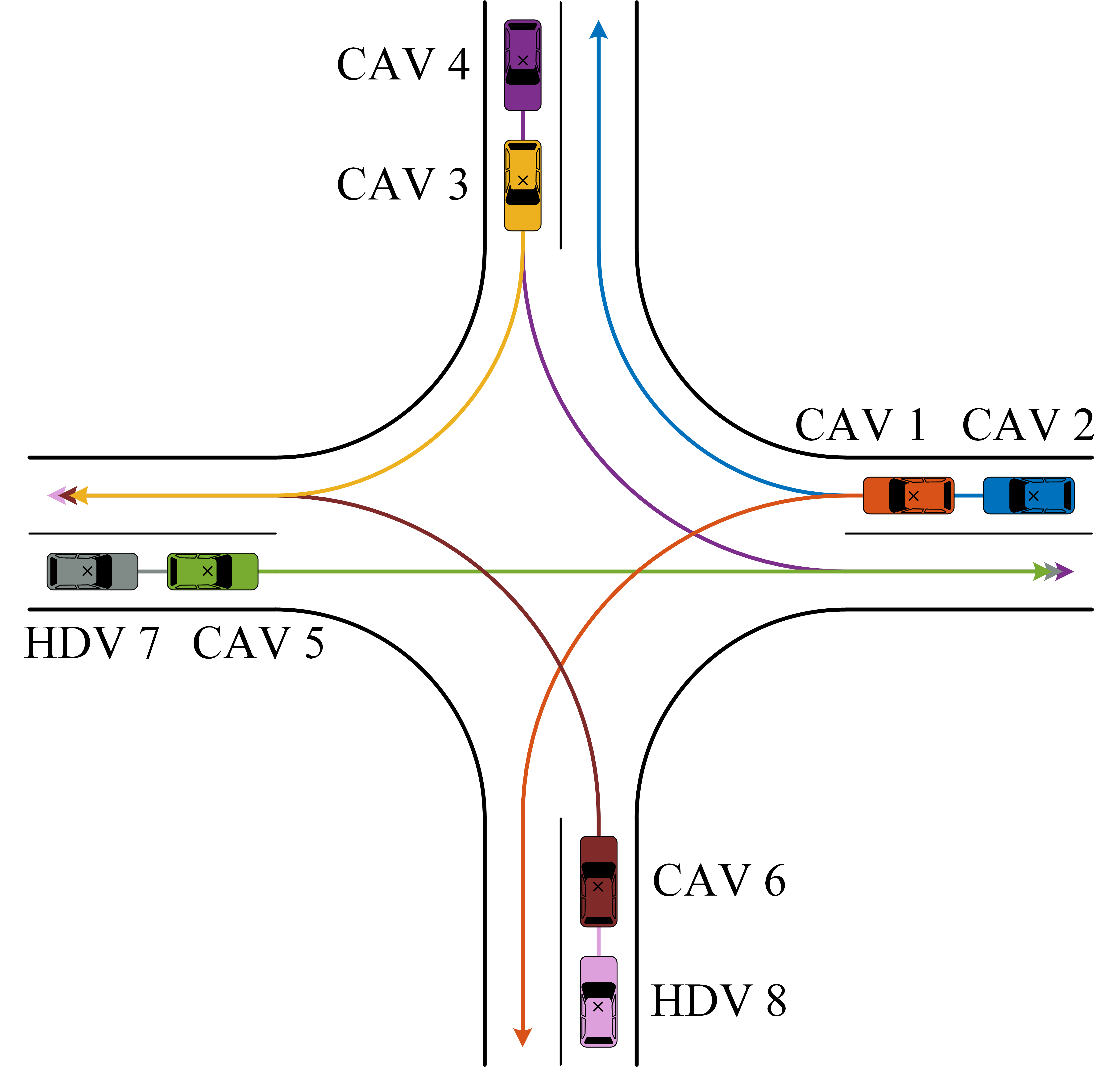}}
\caption{Scenarios investigated in the case study.}
\label{f5}
\end{figure}

In this section, the proposed MPC is validated in the scenarios shown in Fig. \ref{f5}, where the lane width is $4\,\mathrm m$, the central area is $30\,\mathrm m$ across, and the control boundary is a circle with a radius of $90\,\mathrm m$, whose center coincides with the center of the central area. The path speed limit is $v_l^\mathrm{lim}(p)\equiv50\,\mathrm{km/h}$ for all $l\in\mathcal L$. For each CAV $i\in\mathcal M$, the longitudinal acceleration limits are $a_i^{\min}=-3.5\,\mathrm{m/s^2}$ and $a_i^{\max}=2\,\mathrm{m/s^2}$, the maximum acceptable centripetal acceleration is $a_{i\mathrm c}^{\max}=2\,\mathrm{m/s^2}$, and the maximum speed is
\begin{align}
v_i^{\max}(p)=
\begin{cases}
50\,\mathrm{km/h}, \ &\text{if} \ \kappa_i(p)=0,\\	
21.0\,\mathrm{km/h}, \ &\text{if} \ \kappa_i(p)=0.0588,\\
18.4\,\mathrm{km/h}, \ &\text{if} \ \kappa_i(p)=0.0769,
\end{cases}
\end{align}
where the curvature $\kappa_i(p)=0$ corresponds to going straight, $\kappa_i(p)=0.0588$ corresponds to turning left, and $\kappa_i(p)=0.0769$ corresponds to turning right. The desired time gap is set to $t_j^\mathrm{des}=1.1\,\mathrm s$ for all $j\in\mathcal N$. The distance sampling interval is chosen to be $\Delta p=1\,\mathrm m$. The time sampling period is chosen to be $\Delta t=0.5\,\mathrm s$. 

The two cost functions presented in Section \ref{IV-B}, speed tracking and travel time reduction, are used in the case study with the weights 
\begin{equation}
q_{i1}=\frac{w_{i1}\Delta p}{z_{i\mathrm m}^3}, \ r_i=\frac{w_{i2}\Delta p}{z_{i\mathrm m}^5}, \ e_i=\frac{w_{i3}}{\Delta pz_{i\mathrm m}^7}, \ q_{i2}=500,
\end{equation}
where $w_{i1}=1$, $w_{i2}=1$, and $w_{i3}=0.5$ are the weights corresponding to the temporal-domain problem \eqref{10}, as detailed in \cite{murgovski2015convex}, and $z_{i\mathrm m}$ is the mean value of the linearization lethargy $z_{i,\mathrm{lin}}(k)$. In addition, the weight for penalizing the slack variables is $q_\mathrm s=10000$.

The proposed MPC is implemented in MATLAB and the coordinated control problem \eqref{30} is solved by applying RTI and using the optimization tool CasADi. All simulations are performed on a laptop with an Intel Core i9-13900HX CPU at 2.20 GHz and 32 GB RAM.

The simulation results are divided into three parts: the effect of motion uncertainty, the comparison of cost functions, and the analysis of computational effort and optimality.

\begin{table}[htbp]
\centering
\setlength{\tabcolsep}{5pt}
\caption{Specific Vehicle Parameters for Scenario 1 and Scenario 2}
\label{t1}
\begin{tabular}{ll}
\toprule[1pt]
\multicolumn{1}{c}{Description} & \multicolumn{1}{c}{Value}\\
\midrule
Initial position (m) & $\begin{bmatrix} 10 & 15 & 20 & 25 \end{bmatrix}$\\[+1mm]
& $\begin{bmatrix} 25 & 10 & 30 & 0 & 20 & 30 & 5 & 15 \end{bmatrix}$\\[+1mm]
Initial speed (km/h) & $\begin{bmatrix} 40 & 42 & 44 & 46 \end{bmatrix}$\\[+1mm]
& $\begin{bmatrix} 34 & 32 & 42 & 40 & 38 & 46 & 36 & 44 \end{bmatrix}$\\[+1mm]
Initial acceleration ($\text{m/s}^2$) & $\begin{bmatrix} 0 & 0 & 0 & \text{-} \end{bmatrix}$\\[+1mm]
& $\begin{bmatrix} 0 & 0 & 0 & 0 & 0 & 0 & \text{-} & \text{-} \end{bmatrix}$\\[+1mm]
Reference speed (km/h) & $\begin{bmatrix} 40 & 42 & 44 & \text{-} \end{bmatrix}$\\[+1mm]
& $\begin{bmatrix} 50 & 50 & 50 & 50 & 50 & 50 & \text{-} & \text{-} \end{bmatrix}$\\[+1mm]
Crossing order & $\begin{bmatrix} 4 & 2 & 3 & 1 \end{bmatrix}$\\[+1mm]
& $\begin{bmatrix} 3 & 6 & 8 & 1 & 2 & 5 & 7 & 4 \end{bmatrix}$\\
\bottomrule[1pt]
\end{tabular}
\end{table}

\subsection{Effect of Motion Uncertainty}

In this part, the proposed algorithm is applied to Scenario 1, a typical mixed-traffic unsignalized intersection scenario containing both side and rear-end conflicts between CAVs and between CAVs and HDVs, as depicted in Fig. \ref{f5}(a). The exact path of HDV 4 turning left is unknown and is assumed to be contained within the uncertainty set shown in the shaded area. To investigate the effect of motion uncertainty, 100 cases with varying paths for HDV 4 are designed and the same hypothetical speed trajectory is assigned to HDV 4 in all cases for comparison. The initial conditions of the vehicles are listed in Table \ref{t1}. The speed tracking cost is chosen as the cost function.

\begin{figure}[htbp]
\centering
\includegraphics[width=8.6cm,height=6.45cm]{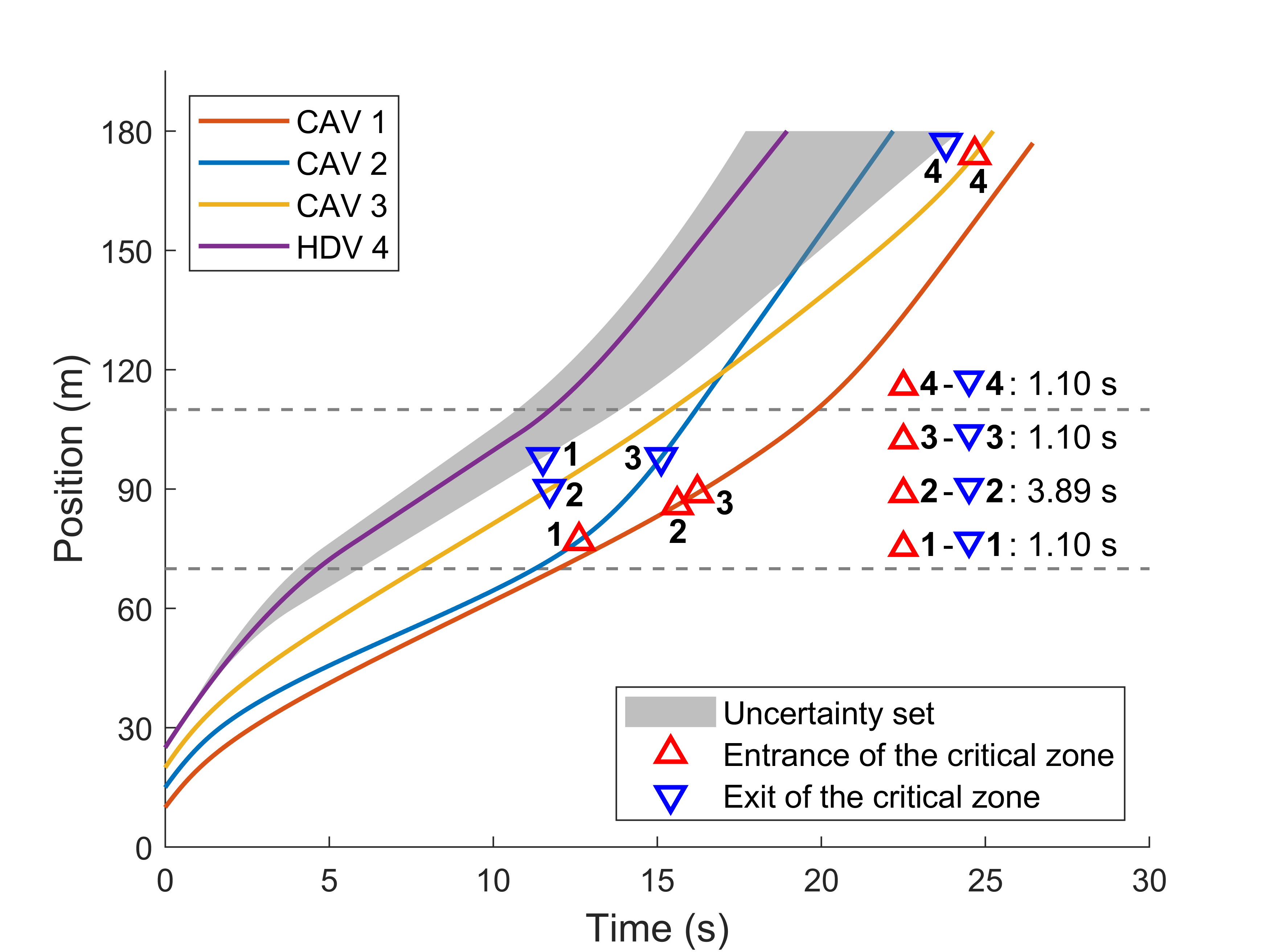}
\caption{Solution of the first MPC iteration for Scenario 1. The purple line indicates the actual trajectory of HDV 4, which is contained within the uncertainty set depicted by the shaded area. For each vehicle pair at risk of collision, the critical zone corresponding to its minimum time gap is labeled by a pair of triangles with the same number.}
\label{f6}
\end{figure}

\begin{figure}[htbp]
\centering
\subfloat{\includegraphics[width=8.6cm,height=6.45cm]{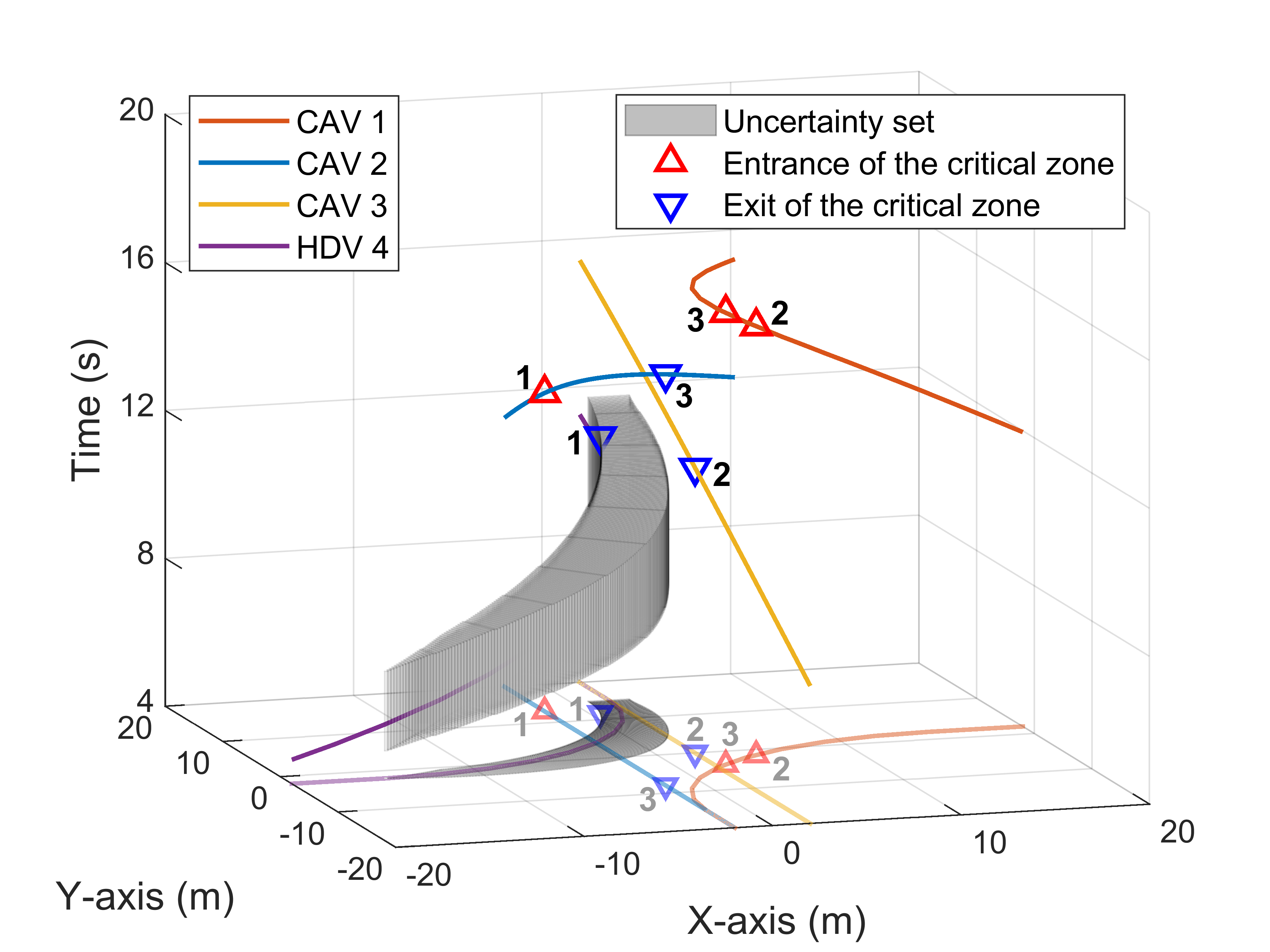}}
\caption{Local 3D spatial-temporal diagram derived from the first MPC iteration for Scenario 1, corresponding to the portion between the dashed lines in Fig. \ref{f6}. The shaded space shows the motion uncertainty of HDV 4, where the X-Y projection plane reflects the path uncertainty and the height reflects the speed uncertainty. The trajectories of all vehicles do not intersect and maintain safety margins, indicating that the solution is collision-free even in the worst case.}
\label{f7}
\end{figure}

The solution of the first MPC iteration, which is the same for all 100 cases, is plotted in Fig. \ref{f6}. The shaded area reflects the speed uncertainty of HDV 4, within which the actual trajectory of HDV 4 is contained, and the upper and lower boundaries are the predicted trajectories with the minimum and maximum travel times, respectively. Since HDV 4 is designated to pass first, the predicted trajectory with the maximum travel time is adopted for robust collision avoidance. For each vehicle pair at risk of collision, the critical zone corresponding to its minimum time gap is labeled by a pair of triangles with the same number. The physical meaning of the time gap for a vehicle pair is the time difference between the follower entering the critical zone and the leader exiting it. It turns out that all the minimum time gaps are equal to or greater than the desired value of $1.1\,\mathrm s$, indicating that the solution fulfills the unrelaxed collision avoidance constraints even in the worst case. Indeed, this property is also present in each subsequent MPC iteration. Furthermore, for visual presentation, the spatial-temporal diagram between the dashed lines in Fig. \ref{f6} is reproduced in Fig. \ref{f7} in three dimensions, from which it is evident that the trajectories of all vehicles do not intersect and maintain safety margins.

\begin{figure}[htbp]
\centering
\includegraphics[width=8.6cm,height=9.83cm]{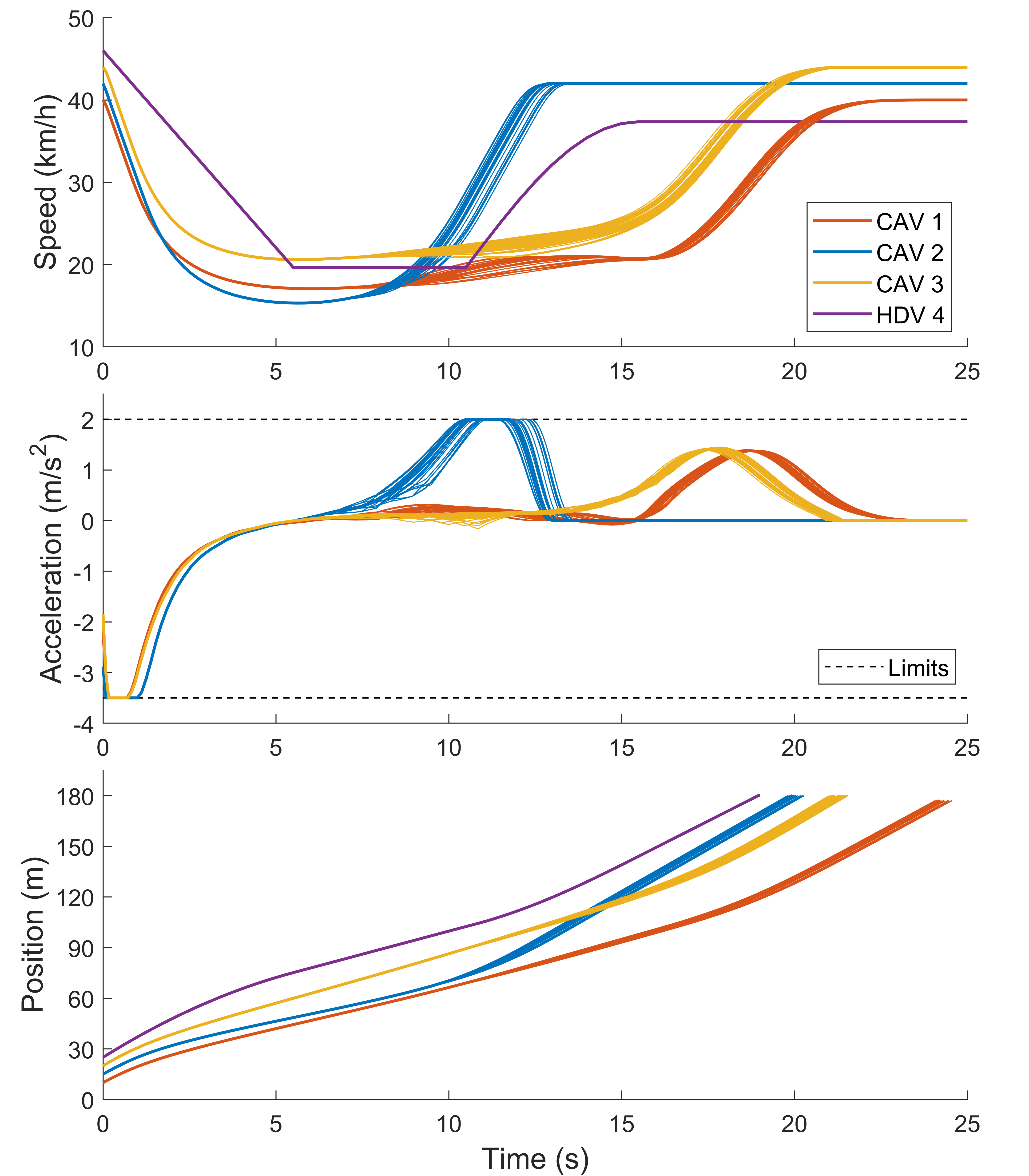}
\caption{Trajectory results of coordinated control for the 100 cases of Scenario 1. Each CAV fulfills the speed and acceleration limits in all cases.}
\label{f8}
\end{figure}

\begin{figure}[htbp]
\centering	\includegraphics[width=8.8cm,height=3.67cm]{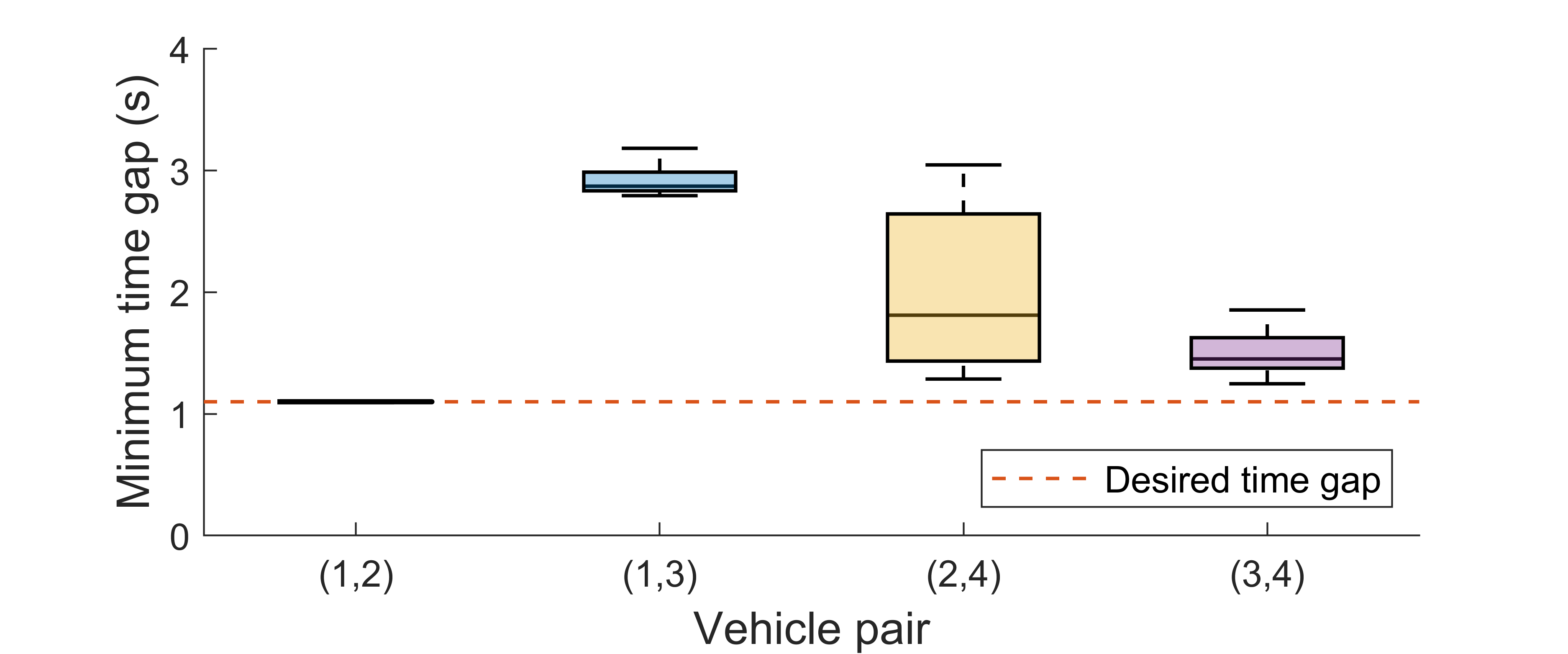}
\caption{Minimum time gaps for vehicle pairs with collision risk for the 100 cases of Scenario 1.}
\label{f9}
\end{figure}

The trajectory results of coordinated control for the 100 cases are presented in Fig. \ref{f8}. HDV 4 is assumed to experience uniform deceleration, uniform speed, variable acceleration, and uniform speed in sequence, as shown by the purple speed profile. The three CAVs first slow down to avoid collision in the central area and later accelerate to their respective reference speeds while ensuring safety. It can be seen that the solutions for the different cases do not differ much and are all smooth. This is because the motion uncertainty of HDV 4 is updated and imposed in each MPC iteration, allowing the CAVs to have enough room for error to adapt to the real situation with only minor control adjustments. To confirm that the solutions are collision-free, box plots are used to show the minimum time gaps for vehicle pairs with collision risk, as depicted in Fig. \ref{f9}. Since all these values are equal to or greater than the desired time gap of $1.1\,\mathrm s$, the solutions for all cases fulfill the unrelaxed collision avoidance constraints.

\begin{figure*}[htbp]
\centering
\subfloat[Case 1: $\text{Time}=9.3\,\mathrm s$.]{\includegraphics[width=5cm,height=4.61cm]{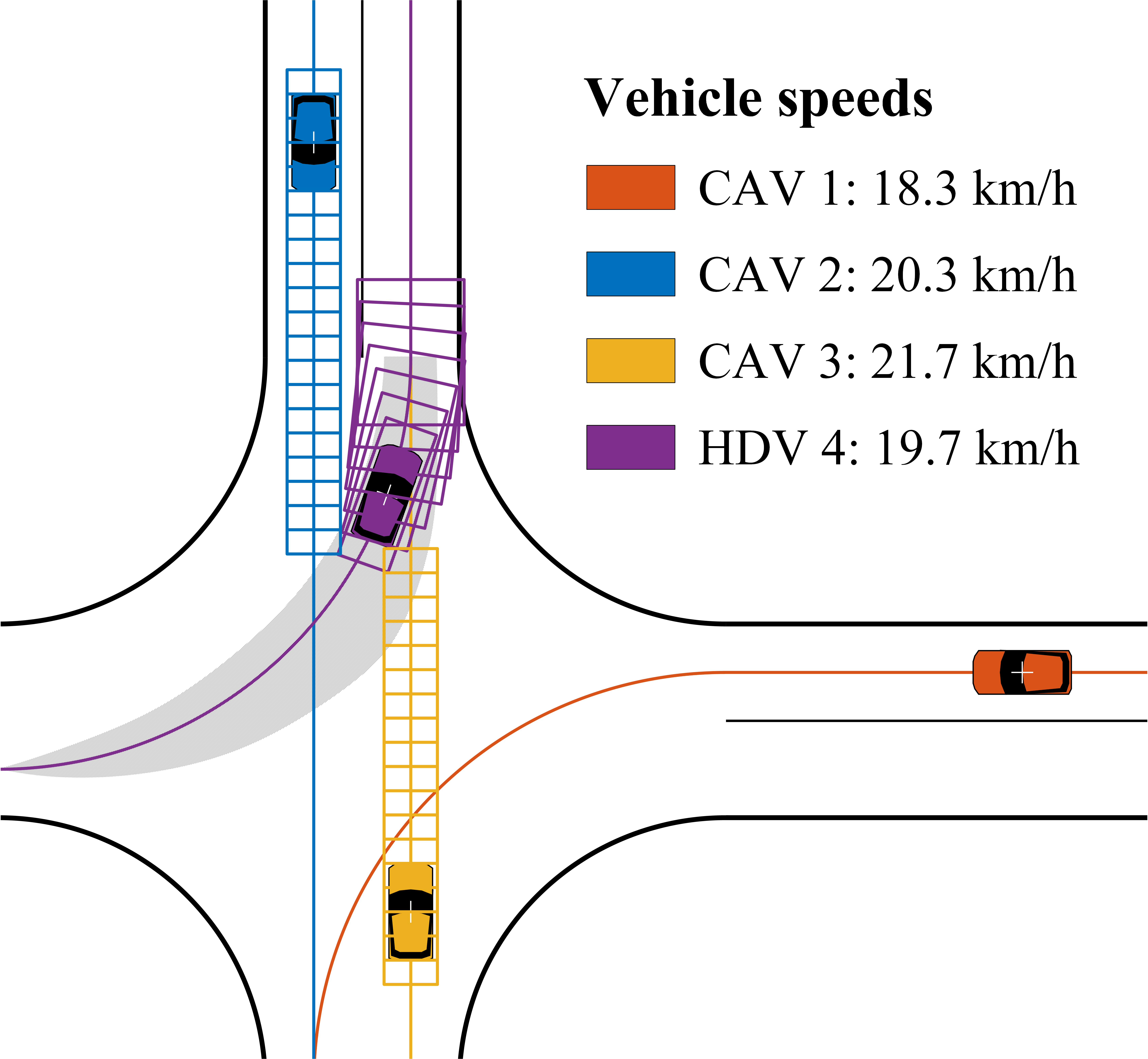}}
\hspace{8mm} 
\subfloat[Case 2: $\text{Time}=9.5\,\mathrm s$.]{\includegraphics[width=5cm,height=4.61cm]{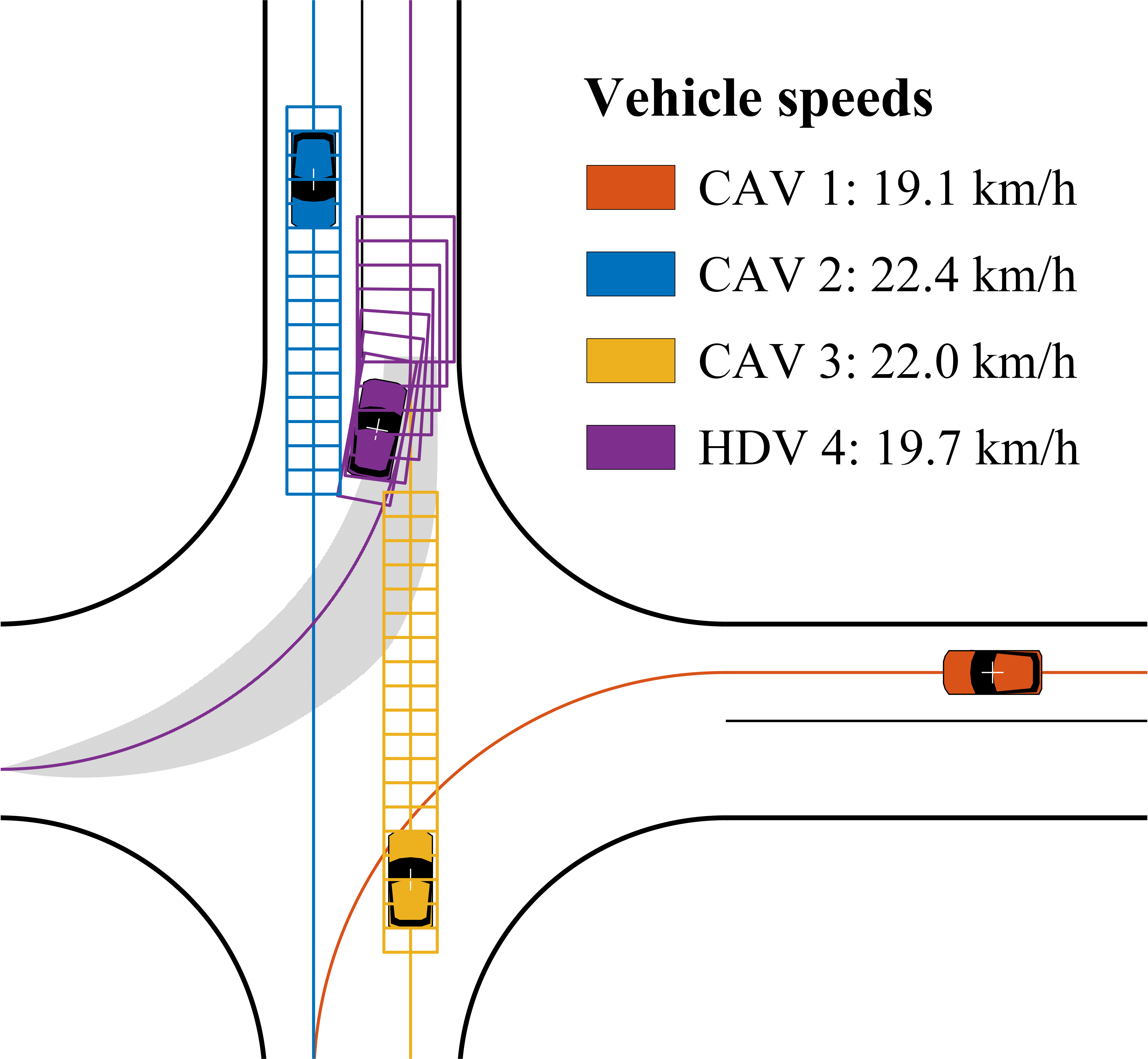}}
\hspace{8mm} 
\subfloat[Case 3: $\text{Time}=8.9\,\mathrm s$.]{\includegraphics[width=5cm,height=4.61cm]{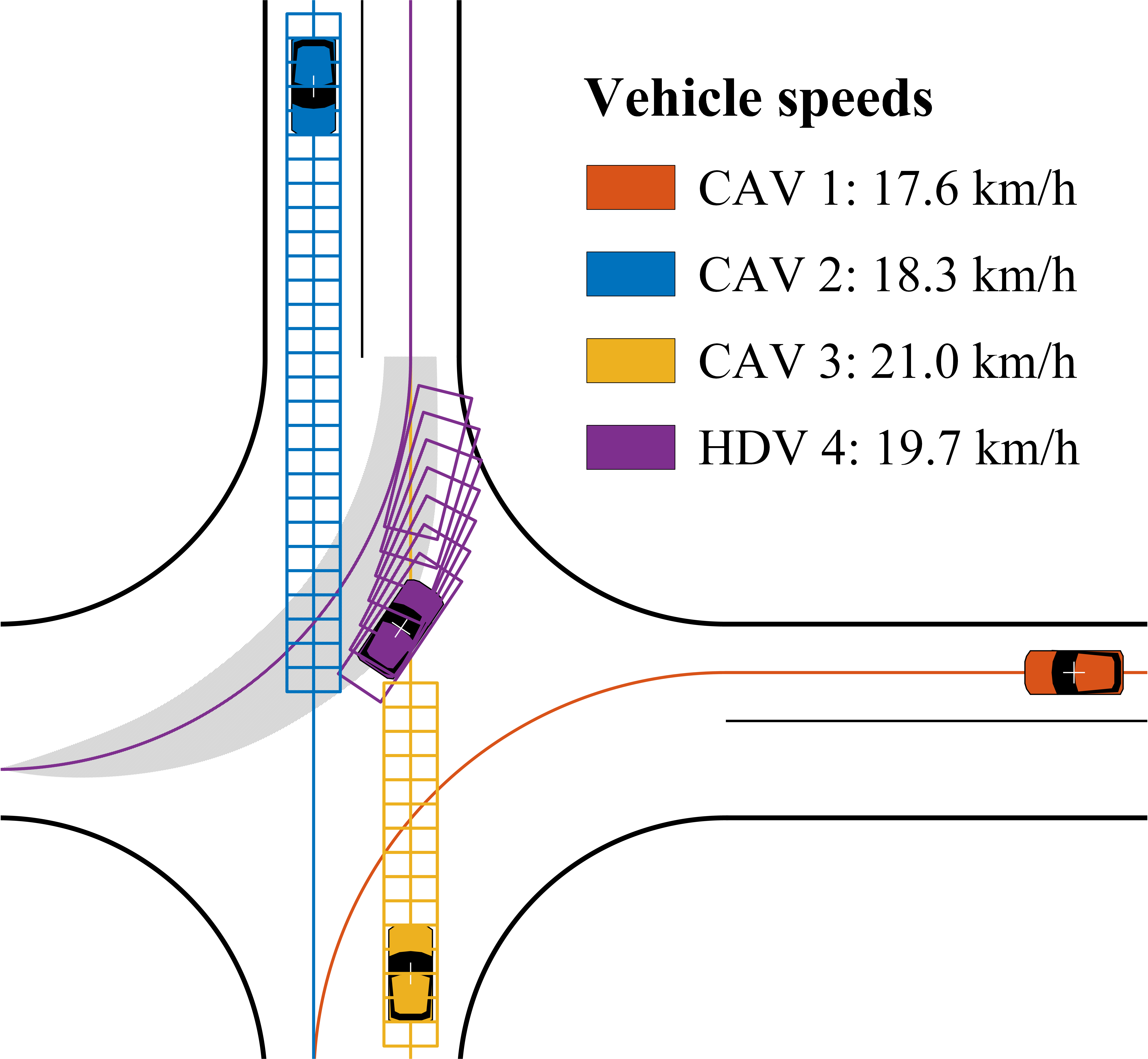}} 
\caption{Snapshots when CAV 2 is about to get rid of the risk of collision with HDV 4 for the three selected cases. The purple bounding box shows the potential path offset of HDV 4 at the spatial sampling point.}
\label{f10}
\end{figure*}

As further shown in Fig. \ref{f9}, the minimum time gap between CAV 1 and CAV 2 is consistently $1.1\,\mathrm s$ across all 100 cases. In contrast, the gap between CAV 2 and HDV 4 ranges from $1.29\,\mathrm s$ to $3.05\,\mathrm s$, and between CAV 3 and HDV 4 from $1.25\,\mathrm s$ to $1.86\,\mathrm s$. This indicates that the motion uncertainty of HDV 4 leads to varying degrees of conservative solutions, somewhat affecting the time efficiency of the CAVs. To visualize this feature, three representative cases are selected for comparison, corresponding in turn to HDV 4 traveling along the reference path, along the counterclockwise boundary of the uncertainty set, and along the clockwise boundary of the uncertainty set. The snapshots when CAV 2 is about to get rid of the risk of collision with HDV 4 are shown in Fig. \ref{f10}, and the minimum time gaps for vehicle pairs with collision risk are listed in Table \ref{t2}. It can be observed that for CAV 2, the solution in Case 2 is the most time-efficient. This is because the collision location in Case 2 is the closest for CAV 2, which constitutes a worst-case scenario, covered by the collision avoidance constraints and playing a decisive role. Accordingly, for the other two cases, more conservative solutions are produced. Nevertheless, these efficiency losses are acceptable given the priority of safety. A similar phenomenon is seen for CAV 3, for which the solution in Case 3 is the most time-efficient, while the solution for CAV 1 is indirectly affected by the motion uncertainty of HDV 4. The conservatism in robust optimization can be reduced by exploiting the distributional characteristics of uncertainty parameters. However, this involves data-driven feature extraction of driving behavior, which is beyond the scope of this paper, and we leave it for future work.
\begin{table}[htbp]
\centering
\setlength{\tabcolsep}{15pt}
\caption{Minimum Time Gaps for Vehicle Pairs With Collision Risk for the Three Selected Cases}
\label{t2}
\begin{tabular}{c|ccc}
\toprule[1pt]
& Case 1 & Case 2 & Case 3\\ 
\midrule
$t_1-t_2$ & $1.10\,\mathrm s$ & $1.10\,\mathrm s$ & $1.10\,\mathrm s$\\[+1mm]
$t_1-t_3$ & $2.89\,\mathrm s$ & $2.80\,\mathrm s$ & $2.88\,\mathrm s$\\[+1mm]
$t_2-t_4$ & $1.91\,\mathrm s$ & $1.29\,\mathrm s$ & $2.92\,\mathrm s$\\[+1mm]
$t_3-t_4$ & $1.64\,\mathrm s$ & $1.86\,\mathrm s$ & $1.31\,\mathrm s$\\
\bottomrule[1pt]
\end{tabular}
\end{table}

Note that when the uncertainty set is not given or fails, the proposed MPC may still work by making initial guesses about the HDV in question, e.g., assuming its speed is constant, assuming it is traveling along a certain feasible path. However, in that case the robustness of collision avoidance is no longer guaranteed and the trajectory smoothness becomes susceptible. In addition, there will inevitably be situations where HDVs do not follow the scheduled crossing order, which requires intent recognition to adjust the crossing order in time. This important consideration is left for future work.

\subsection{Comparison of Cost Functions}

In this part, the proposed algorithm is applied to Scenario 2, where there are two vehicles in each of the four entry lanes, as depicted in Fig. \ref{f5}(b). Scenario 2 comprehensively covers the four types of vehicle conflict situations at unsignalized intersections: crossing, following, merging, and diverging. The motions of HDV 7 and HDV 8 are simulated using the constant time gap car-following law
\begin{equation}
u_i(t)=c_1(d_i(t)-d_0-Tv_i(t))+c_2(v_{i\mathrm l}(t)-v_i(t)),
\end{equation}
where $c_1$ and $c_2$ are feedback gains, $d_i(t)$ is the spacing between HDV $i$ and its leader, $d_0$ is the spacing margin, $T$ is the target time gap, and $v_i(t)$ and $v_{i\mathrm l}(t)$ are the speeds of HDV $i$ and its leader, respectively. Referring to \cite{milanes2014modeling} and \cite{xiao2018unravelling}, $c_1=0.23$, $c_2=0.07$, $d_0=2\,\mathrm m$, and $T=1.4\,\mathrm s$ are taken. The initial conditions of the vehicles are listed in Table \ref{t1}.
\begin{figure*}[htbp]
\centering
\subfloat[]{\includegraphics[width=8.6cm,height=6.66cm]{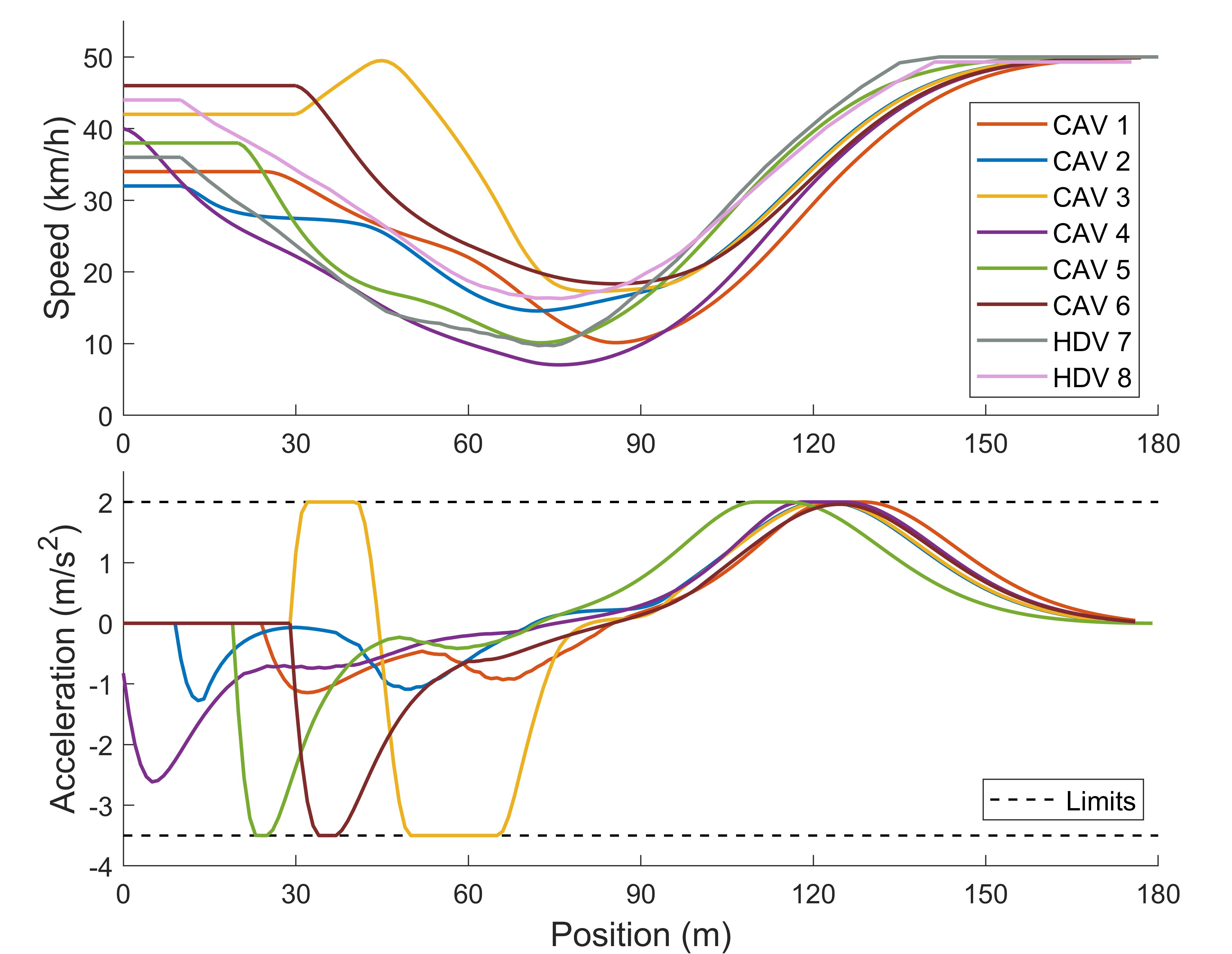}}
\subfloat[]{\includegraphics[width=8.6cm,height=6.66cm]{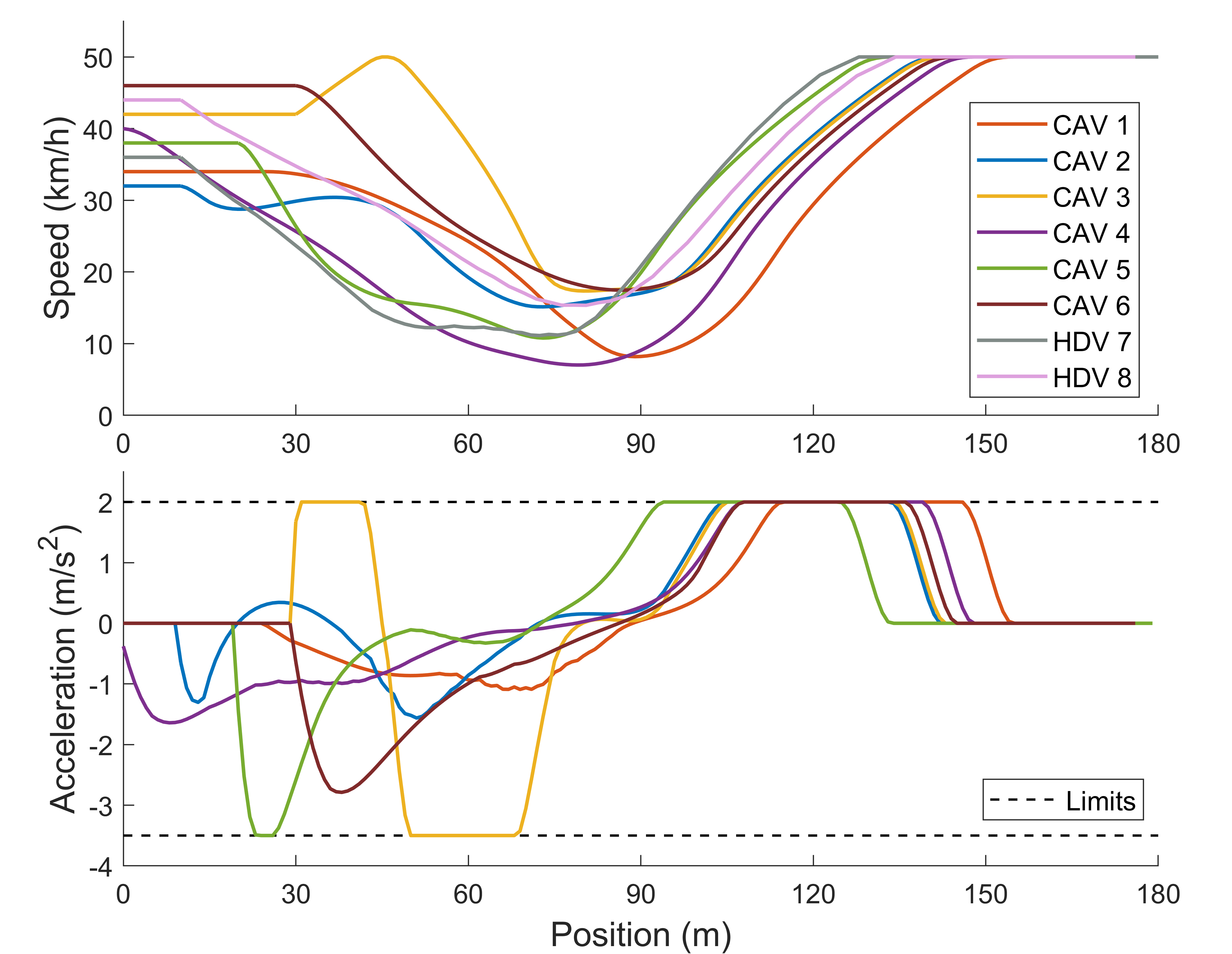}}
\caption{Trajectory results of coordinated control for Scenario 2: (a) using the speed tracking cost; (b) using the travel time reduction cost. All CAVs fulfill the speed and acceleration limits under both costs.}
\label{f11}
\end{figure*}

\begin{figure}[htbp]
\centering
\subfloat{\includegraphics[width=8.2cm,height=0.71cm]{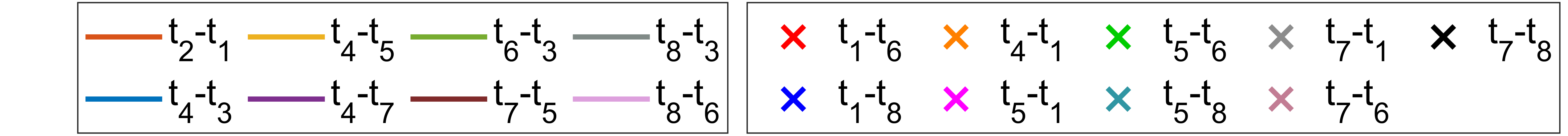}}
\vspace{-3.5mm}
\newline
\subfloat[(a)]{\includegraphics[width=4.4cm,height=4.4cm]{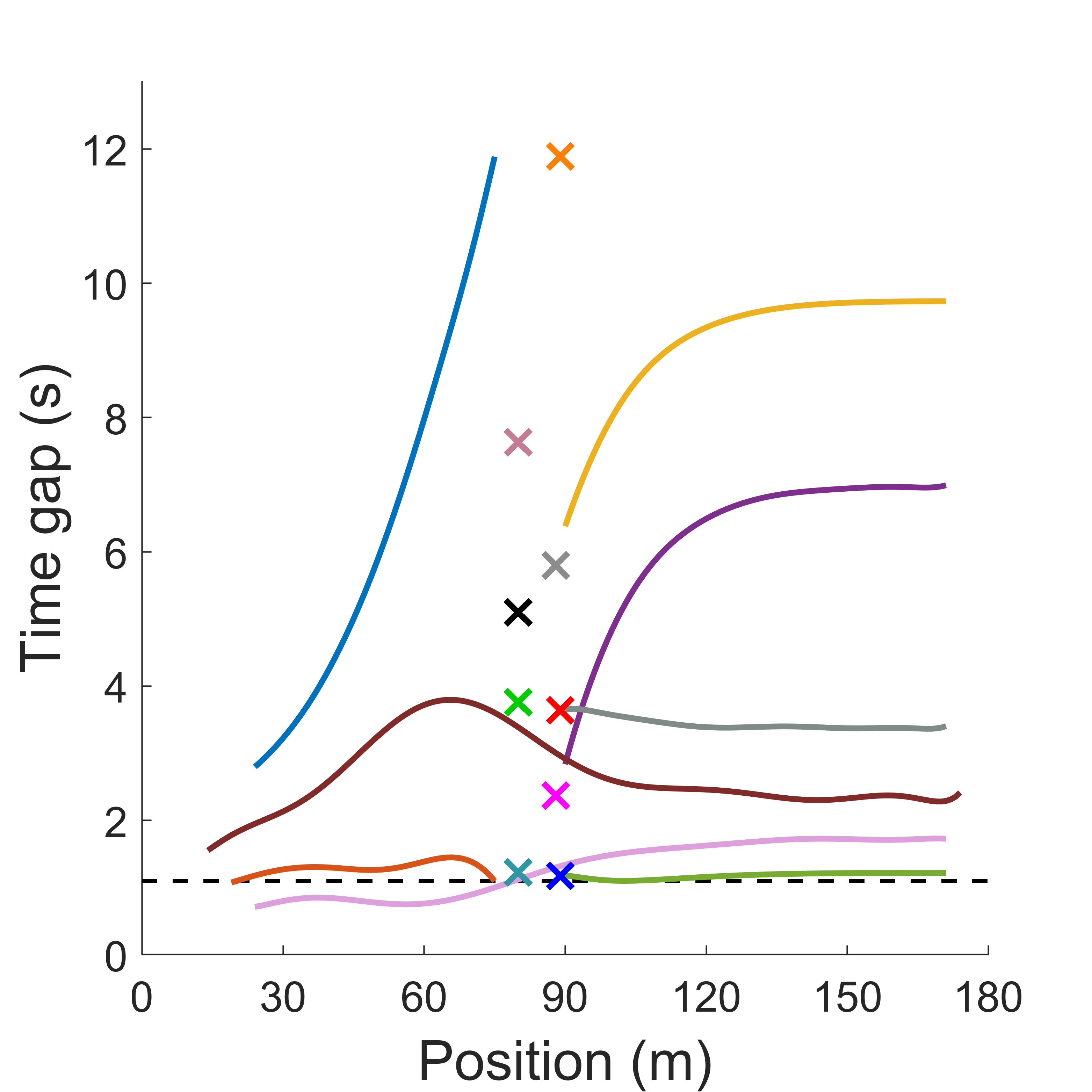}}
\hfill
\subfloat[(b)]{\includegraphics[width=4.4cm,height=4.4cm]{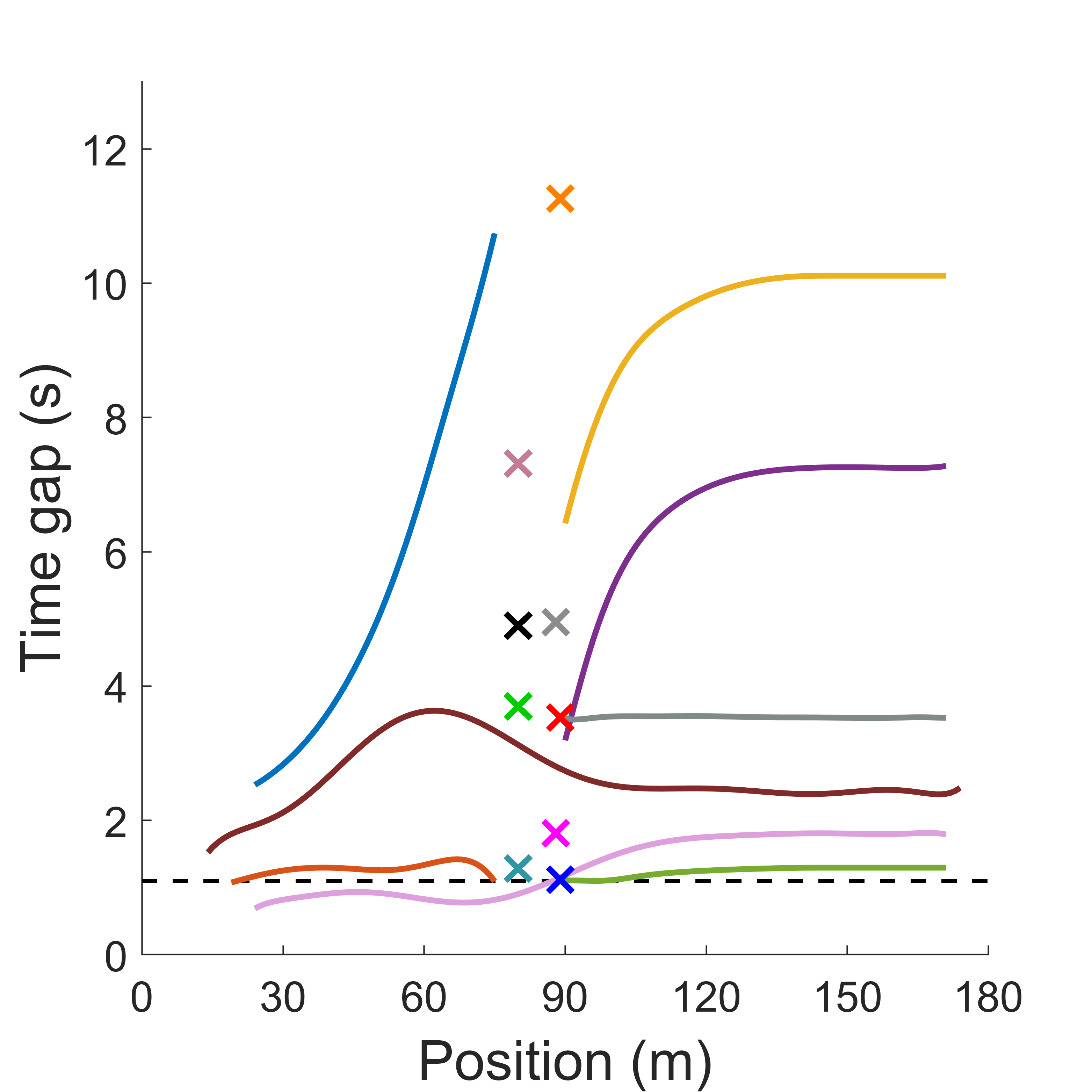}}
\caption{Time gaps for vehicle pairs at risk of collision in Scenario 2: (a) using the speed tracking cost; (b) using the travel time reduction cost. The curves indicate the time gaps for vehicle pairs with overlapping paths, and the crosses indicate the minimum time gaps for vehicle pairs with intersecting paths. The black dashed lines mark the desired time gap of $1.1\,\mathrm s$.}
\label{f12}
\end{figure}

The trajectory results using the speed tracking cost and the travel time reduction cost are presented in Fig. \ref{f11}. To determine that the solutions are collision-free, the time gaps for vehicle pairs at risk of collision are shown in Fig. \ref{f12}, where the curves indicate the time gaps for vehicle pairs with overlapping paths, and the crosses indicate the minimum time gaps for vehicle pairs with intersecting paths. Since all these values are greater than zero, the solutions obtained with both costs are collision-free. Moreover, the time gap between CAV 6 and HDV 8 is observed to be less than the desired value of $1.1\,\mathrm s$ early on (see the pink curves labeled $t_8-t_6$), suggesting that the relaxed collision avoidance constraints are activated to allow for moderate violations of the original hard constraints, thus effectively adapting to the aggressive following behavior of HDVs.
\captionsetup[subfigure]{labelformat=empty}

Although both costs are feasible, there are some differences in the solutions produced. As seen in Fig. \ref{f11}, the travel time reduction cost is more sensitive to speed loss, evident in the fact that the CAVs accelerate faster after passing through the central area and reach the path speed limit of $50\,\mathrm{km/h}$ in a shorter distance, whereas the speed tracking cost yields smoother trajectories at this stage. As expected, the sum of the final travel times of the CAVs is less under the time cost, $137.2\,\mathrm s$, compared to $141.5\,\mathrm s$ under the speed cost, while the final travel times of the last vehicle (CAV 4) are $34.1\,\mathrm s$ and $35.3\,\mathrm s$, respectively. Therefore, the travel time reduction cost motivates CAVs to cross the intersection faster than the speed tracking cost.

\subsection{Analysis of Computational Effort and Optimality}

\begin{figure}[htbp]
\centering
\includegraphics[width=8.6cm,height=6.45cm]{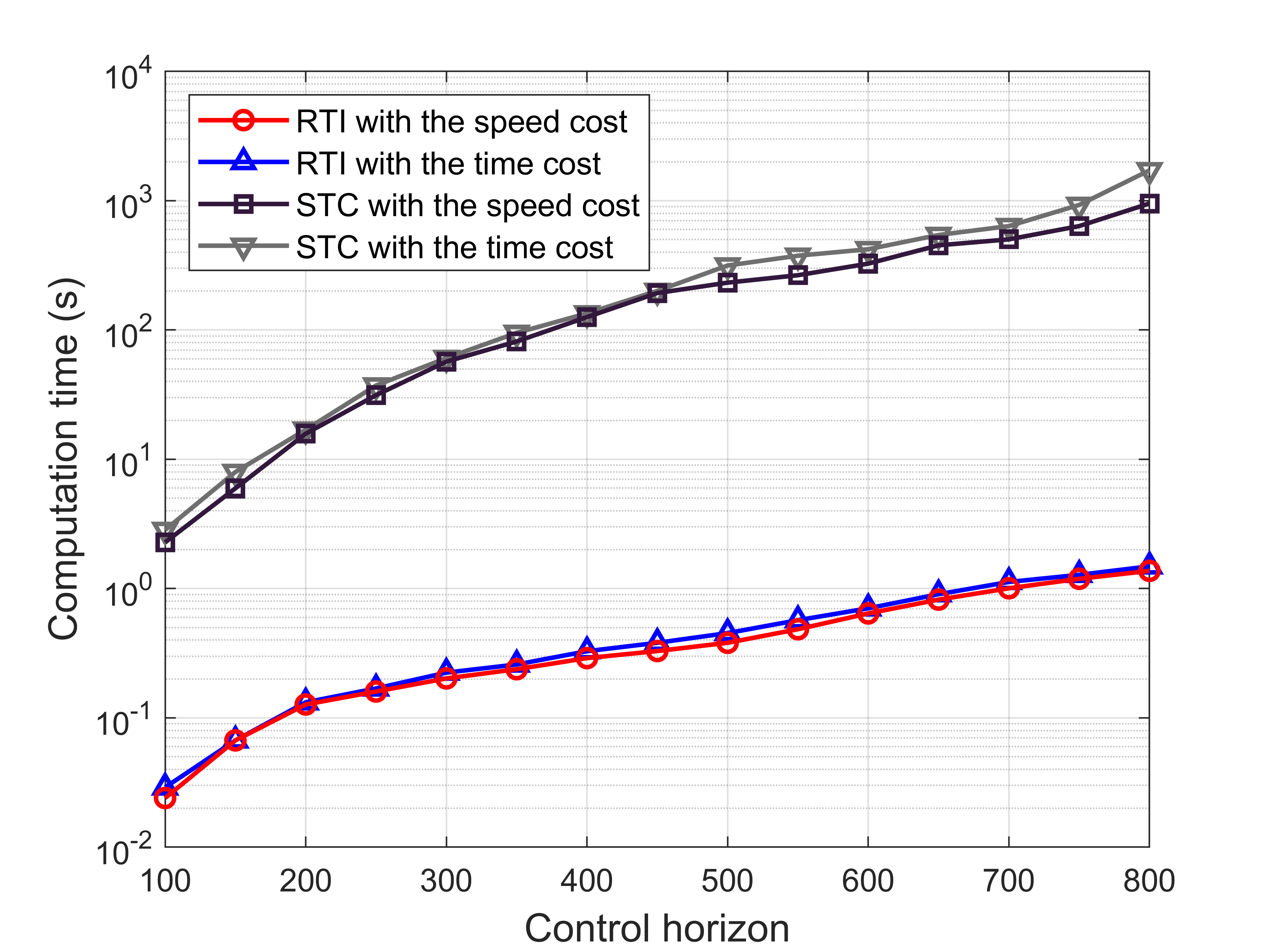}
\caption{Computation time of RTI and STC vs. control horizon.}
\label{f13}
\end{figure}

\begin{figure}[htbp]
\centering
\includegraphics[width=8.6cm,height=6.45cm]{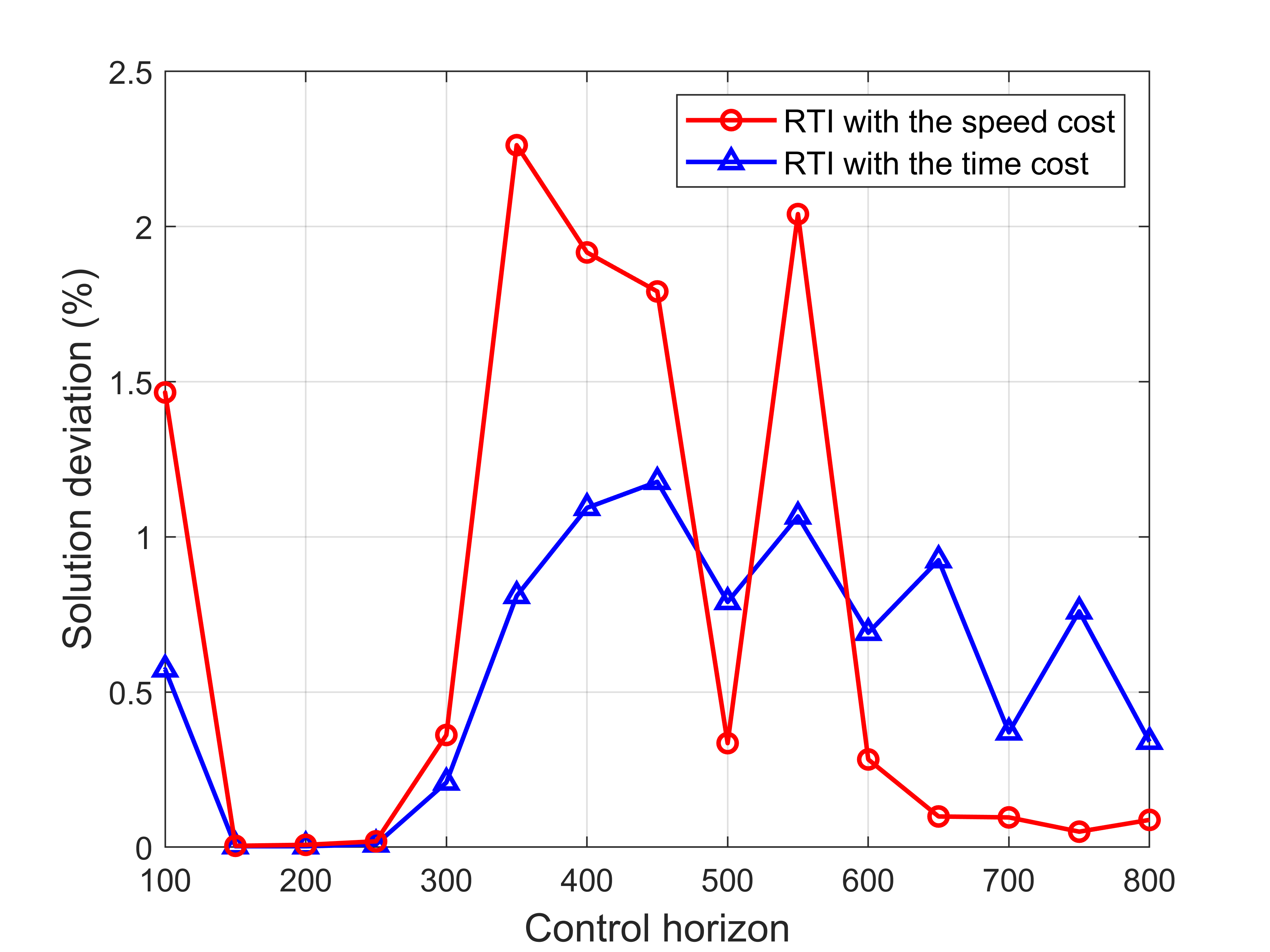}
\caption{Deviation of RTI solution from STC solution vs. control horizon.}
\label{f14}
\end{figure}

In this part, we evaluate the computational effort and optimality of the proposed RTI scheme using Scenario 2. The original NLP \eqref{30} is solved to convergence (this scheme is referred to as STC) for comparison. Specifically, RTI and STC are implemented using the solvers Gurobi and IPOPT, respectively. Fig. \ref{f13} shows the computation time of the two schemes, and Table \ref{t3} summarizes the main performance results. It can be seen that RTI reduces the computation time by orders of magnitude compared to STC, demonstrating good potential for real-time applications. Additionally, the computation time of RTI under the speed tracking cost is slightly lower than that under the travel time reduction cost. Fig. \ref{f14} shows the deviation of the solution obtained by RTI from the solution produced by STC. It turns out that the deviation is less than 2.3\% for different control horizons and cost functions, indicating that RTI yields a solution close to the best known one. Therefore, the proposed RTI scheme achieves a favorable trade-off between computational effort and optimality.
\begin{table*}[htbp]
\centering
\caption{Performance Results of RTI and STC for Control Horizons Ranging From 100 to 800 Samples}
\label{t3}
\begin{tabular}{l|cc|cc}
\toprule[1pt]
& \multicolumn{2}{c|}{STC} & \multicolumn{2}{c}{RTI (one iteration)}\\[+0.5mm]
& Min/Max Comp. time (s) & Min/Max No. of iterations & Min/Max Comp. time (s) & Min/Max Sol. deviation (\%)\\ 
\midrule
Speed tracking cost & $2.278/952.573$ & $24/45$ & $0.024
/1.372$ & $<0.01/2.26$ \\[+1mm]
Travel time reduction cost & $2.841/1719.250$ & $31/91$ & $0.029/1.487$ & $<0.01/1.18$ \\
\bottomrule[1pt]
\end{tabular}
\end{table*}

\begin{figure}[htbp]
\centering
\includegraphics[width=8.8cm,height=6.6cm]{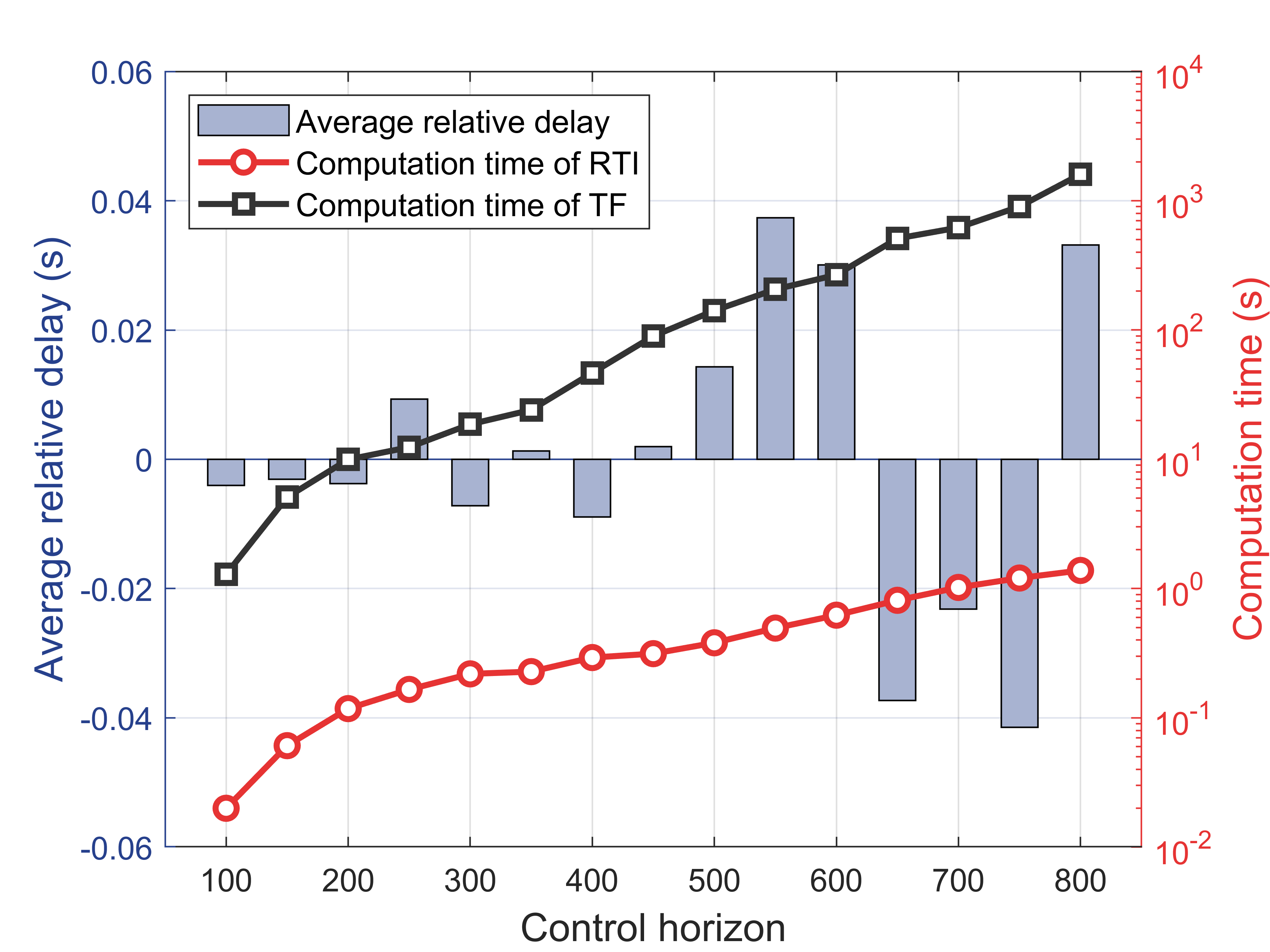}
\caption{Performance comparison between RTI and TF.}
\label{f15}
\end{figure}

Furthermore, we compare the performance of the spatial RTI scheme with the temporal formulation (TF), both of which employ the speed tracking cost function. Referring to \cite{hult2018miqp}, TF is constructed as a mixed-integer quadratic program, and Gurobi is used to solve it. The average CAV delay achieved by RTI relative to TF is calculated to evaluate solution quality. As shown in Fig. \ref{f15}, the absolute value of the average relative delay is less than $0.042\,\mathrm s$, suggesting that the two approaches perform similarly in terms of solution quality. However, as the control horizon extends, TF experiences an explosive increase in computation time, rendering it impractical. This limitation is also shared by STC. In contrast, RTI not only has significantly lower computation time (0.09\% to 1.55\% of TF) but also exhibits a much more moderate growth rate.

It is inevitable that increasing the number of CAVs extends the control horizon, ultimately causing the proposed algorithm to encounter real-time bottlenecks. An effective way to decrease the computational effort is dynamic sampling, i.e., the distance sampling interval increases with vehicle speed, thereby reducing the number of samples in the control horizon. For example, with the distance sampling interval  for each CAV $i\in\mathcal M$ set to $\Delta p=\max\{1\mathrm m,0.2\,\mathrm s/z_{i\mathrm m}\}$ and the travel time reduction cost adopted, Table \ref{t4} presents the RTI computation time statistics for different CAV counts, where each fixed-CAV-count simulation is repeated 100 times with randomized initial conditions. On the other hand, since the actual computation time also depends on hardware and software, the computational efficiency can be further improved by implementing RTI on a dedicated platform, especially given that the formulated subproblem is a convex QP. Therefore, the proposed algorithm is considered to have promising potential for real-time applications in limited-scale scenarios.
\begin{table}[htbp]
\centering
\setlength{\tabcolsep}{12pt}
\caption{Computation Time Statistics for RTI With Dynamic Sampling}
\label{t4}
\begin{tabular}{c|ccc}
\toprule[1pt]
No. of CAVs & 5 & 10 & 15\\ 
\midrule
Avg Comp. time (s) & 0.296 & 1.334 & 1.848\\[+1mm]
Min Comp. time (s) & 0.249 & 1.205 & 1.760\\[+1mm]
Max Comp. time (s) & 0.327 & 1.421 & 1.915\\
\bottomrule[1pt]
\end{tabular}
\end{table}

\section{Conclusion and Future Work}\label{VI}

This paper proposes a coordinated centralized MPC for trajectory planning of CAVs at an unsignalized intersection in mixed traffic. By sampling in distance and using an exact change of variables, the coordinated control problem is formulated in the spatial domain as an NLP. This approach provides simplicity in terms of: 1) handling vehicle crossing, following, merging, and diverging conflicts with unified linear collision avoidance constraints; 2) handling spatially varying speed limits with linear state constraints; and 3) requiring no additional steps to anchor the control horizon. The motion uncertainty of HDVs is modeled at both path and speed levels, based on which the robustness of collision avoidance is ensured in both spatial and temporal dimensions. The subproblem of the NLP is formulated as a convex QP using the inner approximation of the search space, thus enabling the application of RTI for efficient implementation of MPC. The efficacy, robustness, and potential for real-time applications of the proposed methods are demonstrated through simulation case studies. The results show that the proposed control scheme provides collision-free and smooth trajectories with state and control constraints satisfied, and achieves a favorable trade-off between computational effort and optimality. It is worth noting that the proposed methods are applicable to a wider range of right-of-way conflict scenarios, such as roundabouts and ramp merging/diverging.

Future work includes crossing scheduling for unsignalized intersections in mixed traffic. One challenge is to cope with the intent uncertainty of HDVs, determine the right-of-way priority of CAVs with respect to HDVs, and derive a feasible scheduling decision space. Another challenge is to balance the overall computational effort and optimality to achieve efficient cooperative driving that integrates crossing scheduling and trajectory planning. In addition, investigating distributed MPC to enhance the real-time performance of coordinated control is also a research direction worth exploring.

\appendices

\section{Path Uncertainty Set Modeling}\label{A}

For each HDV $i\in\mathcal N\setminus\mathcal M$, its path uncertainty set is written as $\mathscr U_i(\tilde p)=\left[\xi_i^{\min}(\tilde p),\xi_i^{\max}(\tilde p)\right]$, where $\tilde p\in[\tilde p_{i0},\tilde p_{i\mathrm f}]$. The bounds of the path uncertainty set are modeled as 
\begin{equation}
\begin{split}
&\xi_i^{\min}(\tilde p)=\max\left\{\xi_{i\mathrm r}^{\min}(\tilde p),\min\bar\xi_i(\tilde p)\right\},\\
&\xi_i^{\max}(\tilde p)=\min\left\{\xi_{i\mathrm r}^{\max}(\tilde p),\max\bar\xi_i(\tilde p)\right\},
\end{split}
\end{equation}
where the physical limits, $\xi_{i\mathrm r}^{\min}(\tilde p)\leq0$ and $\xi_{i\mathrm r}^{\max}(\tilde p)\geq0$, are preset based on the road geometry, and
\begin{equation}
\begin{split}
\bar\xi_i(\tilde p)=\xi_i(\tilde p_{i0})+\int_{\tilde p_{i0}}^{\tilde p}\tan\theta_i(\tilde s)\mathrm d\tilde s,\\
\theta_i(\tilde p)\in\left[\theta_i^{\min}(\tilde p),\theta_i^{\max}(\tilde p)\right].
\end{split}
\end{equation}
Here, $\xi_i(\tilde p_{i0})$ is the initial offset of HDV $i$, $\tilde s$ is the integral variable corresponding to the projection variable $\tilde p$, and $\theta_i(\tilde p)$ is the yaw angle of HDV $i$ with respect to the reference path, with the limits $\theta_i^{\min}(\tilde p)\leq0$ and $\theta_i^{\max}(\tilde p)\geq0$.

Given the path uncertainty set, any possible path of HDV $i$ can be described by the parametric equations
\begin{equation}
\begin{split}
&x_i(\tilde p,\xi_i(\tilde p))=x_{i\mathrm r}(\tilde p)-\xi_i(\tilde p)\cdot\sin\psi_{i\mathrm r}(\tilde p),\\
&y_i(\tilde p,\xi_i(\tilde p))=y_{i\mathrm r}(\tilde p)+\xi_i(\tilde p)\cdot\cos\psi_{i\mathrm r}(\tilde p),
\end{split}
\end{equation}
where $\tilde p\in[\tilde p_{i0},\tilde p_{i\mathrm f}]$, $\xi_i(\tilde p)\in\mathscr U_i(\tilde p)$, and $x_i(\tilde p,\xi_i(\tilde p))$, $y_i(\tilde p,\xi_i(\tilde p))$ are the global coordinates of HDV $i$ with projection $\tilde p$ and offset $\xi_i(\tilde p)$.

\bibliographystyle{IEEEtran}
\bibliography{IEEEabrv,mybib}

\begin{IEEEbiography}[{\includegraphics[width=1in,height=1.25in,clip,keepaspectratio]{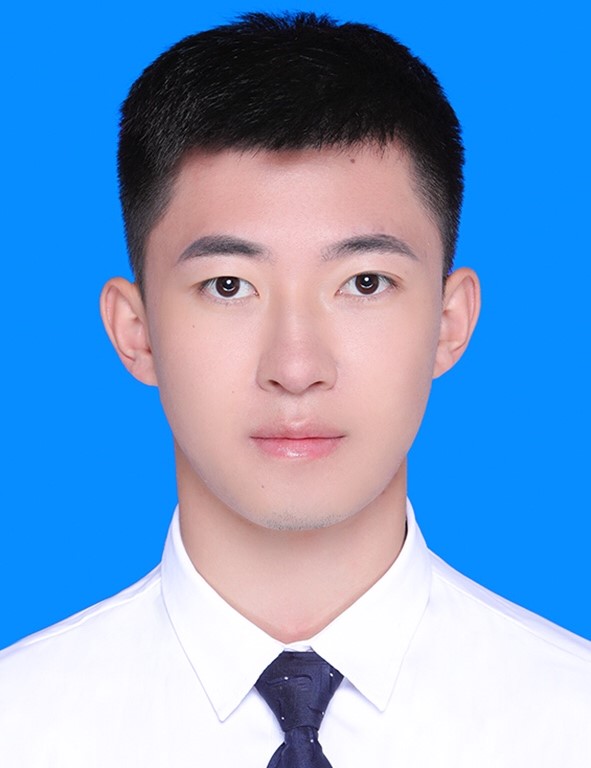}}]{Tong Zhao} received the B.S. degree from Beijing Jiaotong University, Beijing, China, in 2018, where he is currently pursuing the Ph.D. degree with the School of Automation and Intelligence. His research interests include modeling, optimization, and optimal control, with a specific emphasis on connected and automated vehicles.
\end{IEEEbiography}
\begin{IEEEbiography}[{\includegraphics[width=1in,height=1.25in,clip,keepaspectratio]{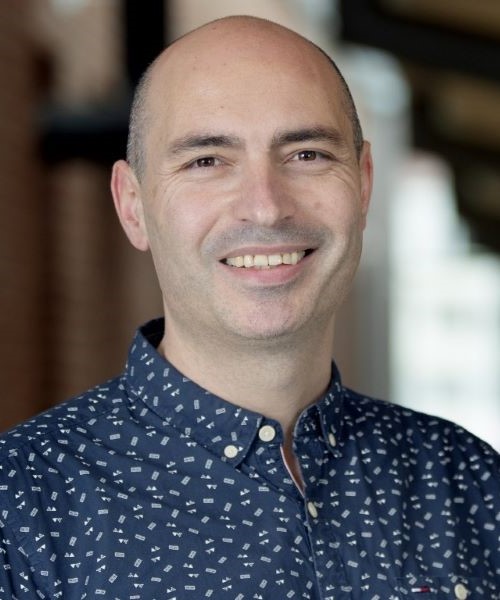}}]{Nikolce Murgovski} received the M.S. degree in Software Engineering from University West, Trollhättan, Sweden, in 2007, and the M.S. degree in Applied Physics and the Ph.D. degree in Systems and Control from Chalmers University of Technology, Gothenburg, Sweden, in 2007 and 2012, respectively. He is currently a Professor with the Department of Electrical Engineering, the Division of Systems and Control, Chalmers University of Technology. His research interests are in modelling, optimization, optimal control, and online learning and estimation. His typical research projects are within electromobility, autonomous driving, and automotive active safety.	
\end{IEEEbiography}
\begin{IEEEbiography}[{\includegraphics[width=1in,height=1.25in,clip,keepaspectratio]{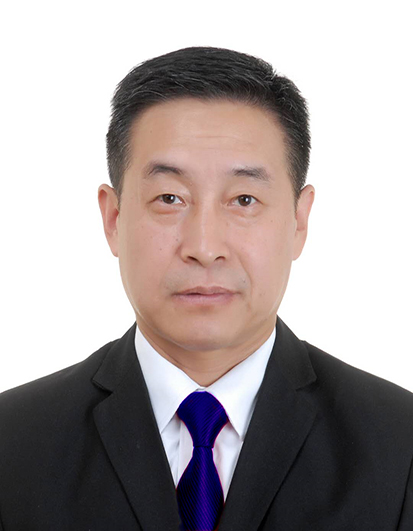}}]{BaiGen Cai} (Senior Member, IEEE) received the B.S., M.S., and Ph.D. degrees from Beijing Jiaotong University, Beijing, China, in 1987, 1990, and 2010, respectively, all in traffic information engineering and control. From 1998 to 1999, he was a Visiting Scholar with Ohio State University, Columbus, U.S.A. He is currently a Professor with the School of Automation and Intelligence, and the Dean at the same School, Beijing Jiaotong University. His research interests include GNSS navigation, intelligent transportation systems, and train control systems.
\end{IEEEbiography}
\begin{IEEEbiography}[{\includegraphics[width=1in,height=1.25in,clip,keepaspectratio]{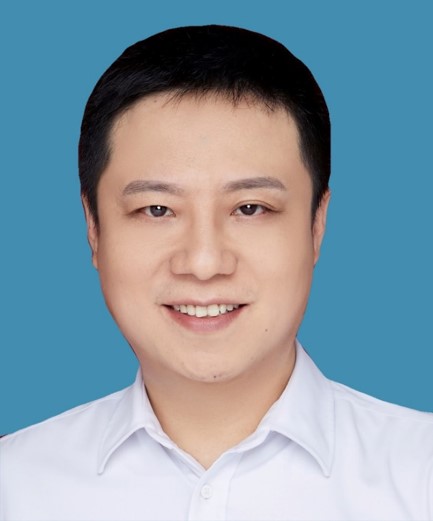}}]{Wei Shangguan}(Member, IEEE) received the B.S., M.S., and Ph.D. degrees from Harbin Engineering University, Harbin, China, in 2002, 2005, and 2008, respectively. From 2013 to 2014, he was an Academic Visitor with University College London, London, U.K. He is currently a Professor with the School of Automation and Intelligence, Beijing Jiaotong University, Beijing, China. His research interests include autonomous intelligence, system modeling, simulation, and testing, intelligent transportation systems, and train control systems.
\end{IEEEbiography}

\end{document}